\documentclass[pdftex,usegraphicx,usenatbib]{mn2e}

\usepackage{graphicx}
\usepackage{natbib}
\usepackage{mathptmx}
\usepackage{amsmath}
\usepackage{amssymb}
\usepackage{wasysym}
\usepackage{url}
\usepackage{nohyperref}
\usepackage{times}
\usepackage{epsfig}
\usepackage{ulem}

\newcommand{\apj}{ApJ}
\newcommand{\mnras}{MNRAS}
\newcommand{\nat}{Nat}

\newcommand{\araa}{ARA\&A}                % "Ann. Rev. Astron. Astrophys."
\newcommand{\aap}{A\&A}                   % "Astron. Astrophys."
\newcommand{\aj}{AJ}                      % "Astron. J."
\newcommand{\apjs}{ApJS}                  % "Astrophys. J. Suppl. Ser."
\newcommand{\apjl}{ApJ}                   % letter at ApJ

\newcommand{\physrep}{Phys. Rep.}

\newcommand{\nar}{NewA Rev.}

\newcommand{\Mpc}{\rm\; Mpc}
\newcommand{\kpc}{\rm\; kpc}
\newcommand{\pc}{\rm\; pc}
\newcommand{\km}{\rm\; km}

\newcommand{\cm}{\rm\; cm}

\newcommand{\pix}{\rm\; pixel}

\newcommand{\ppix}{\hbox{$\pix^{-1}\,$}}

\newcommand{\yr}{\rm\; yr}
\newcommand{\Gyr}{\rm\; Gyr}
\newcommand{\Myr}{\rm\; Myr}
\newcommand{\s}{\rm\; s}

\newcommand{\ks}{\rm\; ks}

\newcommand{\Msun}{\hbox{$\rm\thinspace M_{\odot}$}}

\newcommand{\Msunpyr}{\hbox{$\Msun\yr^{-1}\,$}}
\newcommand{\keV}{\rm\; keV}

\newcommand{\erg}{\rm\; erg}

\newcommand{\ergpcmsqps}{\hbox{$\erg\cm^{-2}\s^{-1}\,$}}

\newcommand{\ergps}{\hbox{$\erg\s^{-1}\,$}}

\newcommand{\keVcmsq}{\hbox{$\keV\cm^{2}\,$}}
\newcommand{\surbri}{\hbox{$\rm\thinspace counts\pcmsq\ps\pasecsq$}}
\newcommand{\expmapcorr}{\hbox{$\rm\thinspace counts\pcmsq\ps\ppix$}}
\newcommand{\kmps}{\hbox{$\km\s^{-1}\,$}}

\newcommand{\Zsun}{\hbox{$\thinspace \mathrm{Z}_{\odot}$}}

\newcommand{\amin}{\rm\; arcmin}

\newcommand{\asec}{\rm\; arcsec}

\newcommand{\kpcpasec}{\hbox{$\kpc\asec^{-1}\,$}}
\newcommand{\psqcm}{\hbox{$\cm^{-2}\,$}}
\newcommand{\pcmsq}{\hbox{$\cm^{-2}\,$}}
\newcommand{\pcmcu}{\hbox{$\cm^{-3}\,$}}

\newcommand{\ps}{\hbox{$\s^{-1}\,$}}

\newcommand{\pasecsq}{\hbox{$\asec^{-2}\,$}}
\begin{document}

\title[Inside the Bondi radius of M87]{The imprints of AGN feedback within a supermassive black hole's sphere of influence}
\author[H.R. Russell et al.]  
    {\parbox[]{7.in}{H.~R. Russell$^{1}$\thanks{E-mail: 
          hrr27@ast.cam.ac.uk}, A.~C. Fabian$^1$, B.~R. McNamara$^{2,3}$, J.~M. Miller$^4$, P.~E.~J. Nulsen$^{5,6}$, J.~M. Piotrowska$^1$, C.~S. Reynolds$^1$\\
    \footnotesize 
    $^1$ Institute of Astronomy, Madingley Road, Cambridge CB3 0HA\\
    $^2$ Department of Physics and Astronomy, University of Waterloo, Waterloo, ON N2L 3G1, Canada\\
    $^3$ Perimeter Institute for Theoretical Physics, Waterloo, Canada\\ 
    $^4$ Department of Astronomy, University of Michigan, 1085 South University Avenue, Ann Arbor, MI 48109-1107, USA \\
    $^5$ Harvard-Smithsonian Center for Astrophysics, 60 Garden Street, Cambridge, MA 02138, USA\\
    $^6$ ICRAR, University of Western Australia, 35 Stirling Hwy, Crawley, WA 6009, Australia\\
  }
}

\maketitle

% Rapid cooling of the hot gas atmosphere within a supermassive black hole's sphere of influence
% On proposal: Andy Fabian, Brian McNamara, Paul Nulsen, Chris Reynolds.  Add in Jon Miller and Joanna Piotrowska

\begin{abstract}
We present a new $300\ks$ \textit{Chandra} observation of M87 that limits pileup to only a few per cent of photon events and maps the hot gas properties closer to the nucleus than has previously been possible.  Within the supermassive black hole's gravitational sphere of influence, the hot gas is multiphase and spans temperatures from $0.2$ to $1\keV$.  The radiative cooling time of the lowest temperature gas drops to only $0.1-0.5\Myr$, which is comparable to its free fall time.  Whilst the temperature structure is remarkably symmetric about the nucleus, the density gradient is steep in sectors to the N and S, with $\rho{\propto}r^{-1.5\pm0.1}$, and significantly shallower along the jet axis to the E, where $\rho{\propto}r^{-0.93\pm0.07}$.  The density structure within the Bondi radius is therefore consistent with steady inflows perpendicular to the jet axis and an outflow directed E along the jet axis.  By putting limits on the radial flow speed, we rule out Bondi accretion on the scale resolved at the Bondi radius.  We show that deprojected spectra extracted within the Bondi radius can be equivalently fit with only a single cooling flow model, where gas cools from $1.5\keV$ down below $0.1\keV$ at a rate of $0.03\Msunpyr$.  For the alternative multi-temperature spectral fits, the emission measures for each temperature component are also consistent with a cooling flow model.  The lowest temperature and most rapidly cooling gas in M87 is therefore located at the smallest radii at $\sim100\pc$ and may form a mini cooling flow.  If this cooling gas has some angular momentum, it will feed into the cold gas disk around the nucleus, which has a radius of $\sim80\pc$ and therefore lies just inside the observed transition in the hot gas structure.
\end{abstract}


\section{Introduction}
\label{sec:intro}

Accretion onto supermassive black holes (SMBHs) powers the intense
radiation observed from distant quasars and spectacular relativistic
jets in radio galaxies that can reach far beyond the host galaxy.  The
interactions of these energetic outbursts with the gas in the host
galaxy are now understood to be a key mechanism in galaxy evolution
that can truncate the galaxy's growth
(\citealt{Croton06,Bower06,Hopkins06}).  By driving out and heating
the surrounding gas, the active galactic nucleus (AGN) can suppress
star formation but also restrict the fuel available for accretion.
Known as AGN feedback, this process has two main modes (for a review
see eg. \citealt{Fabian12}).  The radiatively efficient
quasar-mode operates at high accretion rates, typically above 10\% of
the Eddington rate.  However, the vast majority of SMBHs are accreting
well-below their Eddington limit (\citealt{Ho08,Ho09}).  These black
holes are typically embedded in the substantial hot atmospheres of
their host galaxies but are remarkably faint and must be accreting in
a radiatively inefficient or radio mode (for reviews see
\citealt{Narayan08,Yuan14}).  The fuelling mechanism is therefore
crucial for our understanding of AGN feedback but requires that we
resolve structure on scales within the SMBH's gravitational sphere of
influence.

%The energy produced by the growth of the black hole over its lifetime
%far exceeds the binding energy of the host galaxy.

The classical Bondi model states that the galaxy's hot gas atmosphere
will be accreted by the SMBH if it falls within the Bondi radius,
where the gravitational potential of the SMBH dominates over the
thermal energy of the gas (\citealt{Bondi52}).  The Bondi model assumes spherical accretion onto a point mass and an absorbing inner boundary condition.  The Bondi radius is
given by $r_{\mathrm{B}}=2GM_{\mathrm{BH}}/c^{2}_{\mathrm{s}}$, where
$c_{\mathrm{s}}$ is the sound speed in the hot gas, and is at most a
few arcsec for the nearest and most massive black holes.  The X-ray
emitting hot gas within this region can only currently be resolved in
a handful of systems.  The central galaxy in the Virgo cluster, M87,
has bright thermal gas emission within the Bondi radius, which
dominates over that from stellar sources and LMXBs.  This hot gas may
be fuelling the powerful radio-jet outburst
(\citealt{Bohringer95,Bicknell96,Young02,FormanM8705,Werner10}).
Using early \textit{Chandra} observations of M87, \citet{DiMatteo03}
extracted the gas temperature and density structure at the galaxy
centre and determined the Bondi radius to be $\sim0.15\kpc$ (for a
black hole mass $M_{\mathrm{BH}}=3\times10^{9}\Msun$; \citealt{Ford94,Harms94,Macchetto97}).  Recent gas-dynamical and
stellar-dynamical analyses of M87 estimate a supermassive black hole
mass $M_{\mathrm{BH}}=3-6\times10^{9}\Msun$
(\citealt{Gebhardt11,Walsh13}), which corresponds to a Bondi radius
$r_{\mathrm{B}}=0.12-0.22\kpc$ ($1.5-2.8\asec$).  With deeper
\textit{Chandra} observations, the relatively high count rate from the
hot atmosphere, compared to other possible targets, should produce the
most detailed maps of the gas density and multi-temperature structure
within a Bondi sphere.  However, the hot gas structure within the
Bondi radius of M87 is overwhelmed by the bright nuclear point source.
From 2003 to 2010, the Bondi sphere was also inaccessible with
\textit{Chandra} beneath heavy pileup due to flaring of the jet knot
HST-1, which is only $0.85\asec$ from the nucleus
(\citealt{Harris06,Harris09}).

Although fainter, the hot gas emission within the Bondi radius of Sgr
A* at the centre of the Milky Way and the nearby galaxy NGC\,3115 were
subsequently observed for multiple Ms with \textit{Chandra}
(\citealt{Baganoff03,Wong11,Wang13,Wong14}).  These observations found
shallow gas density profiles inside the Bondi radius with
$\rho{\propto}r^{-1}$.  Numerical simulations of radiatively
inefficient accretion flows (ADAFs,
\citealt{Ichimaru77,Rees82,Narayan94,Narayan08}) show that winds launched from
the hot accretion flow will drive out a large fraction of the
accreting gas and explain the observed shallow density slopes
(eg. \citealt{Stone99,Stone01,Hawley02,Yuan12,Li13,Yuan14}).  Whilst
even a substantially reduced accretion rate provides ample fuel for
these quiescent systems, it is likely insufficient to power the
radio-jet outbursts that are a common feature of radiatively
inefficient accretion flows (eg. \citealt{Rafferty06,Nemmen15}).  These systems
may require a supplementary inflow from cold gas accretion
(\citealt{PizzolatoSoker05,Gaspari13}) or additional power from black
hole spin energy (\citealt{McNamara11,Tchekhovskoy11,Tchekhovskoy12,McKinney12}).

Deep \textit{Chandra} observations of M87 have previously focused on the
large cavities and weak shocks in the hot atmosphere where the radio
jet and expanding lobes have displaced the surrounding medium
(\citealt{Bohringer95,Bicknell96,Young02,FormanM8705,Werner10}).  The
energy input by the jet is replacing the radiative losses from the
galaxy's X-ray atmosphere to keep the gas hot and prevent the
formation of a large-scale cooling flow (for reviews see
\citealt{PetersonFabian06,McNamaraNulsen07}).  The nucleus in M87 is also the target of ongoing
\textit{Chandra} monitoring observations, which revealed a
decline in the jet knot brightness after the peak around 2005
(\citealt{Harris09,Harris11}).  By 2010 the jet knot
brightness had dropped back to the original level observed in 2000,
therefore \citet{Russell15} were able to stack twelve recent monitoring
observations to produce radial temperature and density profiles of the
hot gas within the Bondi radius.  This analysis indicated a multiphase
structure and a gas density profile consistent with
$\rho{\propto}r^{-1}$, although this measurement was potentially
affected by the inner cavity structure.

Here we present a new $300\ks$ short frame time \textit{Chandra}
observation of the centre of M87 that limits pileup of the nuclear
emission to a few per cent.  This allows us to map the detailed
multi-temperature structure within the Bondi radius and study the
response of the gas flow to the most recent AGN outburst on smaller
scales than has previously been possible.  We assume a distance to M87 of
$16.1\Mpc$ (\citealt{Tonry01}) to be consistent with earlier analyses
of the \textit{Chandra} datasets.  This gives a linear scale of
$0.078\kpcpasec$ and $\sim2-5\times10^{5}$ gravitational radii per
arcsec, depending on the black hole mass.  All errors are $1\sigma$
unless otherwise noted.

\begin{table*}
\begin{minipage}{\textwidth}
\caption{Details of the \textit{Chandra} observations used for this analysis and best-fit nuclear spectral model parameters.}
\begin{center}
\begin{tabular}{l c c c c c c}
\hline
Obs. ID & Date & Exposure & Frame time & $n_{\mathrm{H}}$ & $\Gamma$ & Flux ($2-10\keV$)\\
 & & (ks) & (s) & ($10^{22}\psqcm$) & & ($10^{-12}\ergpcmsqps$) \\
\hline
352 & 29/07/2000 & 29.4 & 3.2 & - & - & - \\
1808 & 30/07/2000 & 12.8 & 0.4 & $0.08\pm0.01$ & $2.37\pm0.07$ & $0.70\pm0.04$ \\
11513 - 14974 & 2010 - 2012 & 54.7 & 0.4 & $0.057\pm0.006$ & $2.25\pm0.03$ & $1.30\pm0.03$ \\
18232 & 27/04/2016 & 18.2 & 0.4 & $0.08\pm0.02$ & $2.29\pm0.07$ & $0.70^{+0.04}_{-0.03}$ \\
18233 & 23/02/2016 & 37.2 & 0.4 & $0.03\pm0.01$ & $2.27\pm0.05$ & $0.62\pm0.02$ \\
18781 & 24/02/2016 & 39.5 & 0.4 & $0.04\pm0.01$ & $2.21\pm0.05$ & $0.67\pm0.02$ \\
18782 & 26/02/2016 & 34.1 & 0.4 & $0.04\pm0.02$ & $2.21\pm0.05$ & $0.70\pm0.03$ \\
18783 & 20/04/2016 & 36.1 & 0.4 & $0.05\pm0.02$ & $2.28\pm0.05$ & $0.55\pm0.02$ \\
18836 & 28/04/2016 & 38.8 & 0.4 & $0.08\pm0.02$ & $2.27\pm0.05$ & $0.72\pm0.02$ \\
18837 & 30/04/2016 & 13.7 & 0.4 & $0.04\pm0.03$ & $2.32\pm0.09$ & $0.52^{+0.04}_{-0.03}$ \\
18838 & 28/05/2016 & 56.3 & 0.4 & $0.04\pm0.01$ & $2.30\pm0.05$ & $0.50\pm0.02$ \\
18856 & 12/06/2016 & 24.5 & 0.4 & $0.05\pm0.02$ & $2.30\pm0.07$ & $0.50^{+0.03}_{-0.02}$ \\
18232 - 18856 & 2016 & 298.4 & 0.4 & $0.052\pm0.006$ & $2.27\pm0.02$ & $0.537^{+0.008}_{-0.007}$ \\
\hline
\end{tabular}
\end{center}
\label{tab:obs}
\end{minipage}
\end{table*}

\section{\textit{Chandra} data analysis}
\label{sec:dataanalysis}

The central $1\amin$ region of M87 was observed for a new large
\textit{Chandra} program on ACIS-S to map the properties of the hot
gas atmosphere within the Bondi radius (obs. IDs 18232 to 18856, Table
\ref{tab:obs}).  The total exposure of $298\ks$ was taken with a short
frame time of $0.4\s$ and used a subarray to minimize pileup.  In
these new observations, the nucleus and jet knot HST-1 have declined
in brightness by more than a factor of $\sim2$ below the levels
measured in our earlier analysis (obs. IDs 11513 to 14974,
\citealt{Russell15}).  As contamination of the hot gas emission by the
nuclear and jet knot emission is the key limitation in our analysis,
we have primarily used the new observations to extract the hot gas
properties and thereby probe further into the Bondi sphere.

This analysis also utilized archival observations of M87 from 2000
(obs. ID 1808) and 2010 to 2012 (obs. IDs 11513 to 14974) to test the
subtraction of the nuclear and jet knot emission (section
\ref{sec:subpsf}).  The ACIS-S observation without a subarray in 2000
(obs. ID 352) was also used to determine the gas properties at large
radii for the deprojection analysis (section \ref{sec:spec}).

\subsection{Data reduction}
\label{sec:datared}

\textsc{ciao} version 4.9 and \textsc{caldb} version 4.7.3 provided by
the \textit{Chandra} X-ray Center (\citealt{Fruscione06}) were used to
analyse the new observations.  The level 1 event files were reprocessed to
ensure that a consistent calibration was applied to each dataset
including the latest charge transfer inefficiency correction and gain
adjustments.  No major flare periods were found in background light
curves extracted from each new dataset.  The background light curves
were also compared to ensure that no flares were missed in the shorter
observations.  Two short flares occurred during obs. ID 352 and the
corresponding time periods were excluded from the analysis.  The
datasets used and their final cleaned exposure times are detailed in
Table \ref{tab:obs}.

% Background emission

Due to the proximity of M87, the X-ray emission from the hot
atmosphere can be detected across the full extent of the ACIS
detector.  Therefore, blank sky background observations were required
to subtract the X-ray background emission from each dataset.  Each
appropriate background observation was reprocessed in the same way as
the source files and reprojected to the corresponding sky position.
The exposure time of the background dataset was then scaled so that
the background count rate matched that of the source file in the
$9.5-12\keV$ energy band, which ensured correct normalization of the
particle background.  Images that combined multiple observations
required corresponding total backgrounds where the ratio of the
background exposure time to the source exposure time was the same
(12:1) for every input observation.  We therefore discarded events at
random from each blank sky background observation to reduce the
exposure time and maintain the count rate.

% Point sources

Point sources were detected in a summed hard band image ($3-7\keV$)
that included all the 2016 datasets, confirmed by eye and excluded
from the analysis.  As discussed in \citet{Russell15}, the
emission from the hot atmosphere in M87 dominates over any remaining contaminating flux from
unresolved sources, including low mass X-ray binaries, cataclysmic
variables and coronally active binaries.

% Centering methods used, jet emission

Sub-pixel event repositioning was used to separate the nucleus from
the jet knot HST-1, which is located $0.8\asec$ away
(\citealt{Tsunemi01,Li04,Miller17}).  For each observation, an image
of the nucleus and nearby jet knots was generated using a spatial
resolution 10 times finer than the native resolution of $0.495\asec$.
The centre of each emission peak was then determined with 2D image
fitting in \textsc{sherpa}, which utilized the \textsc{simplex}
optimizer and separate 2D Gaussian components for each source.  All
observations were aligned according to the best-fit positions.
Although the nuclear emission cannot be cleanly disentangled from the
jet knot HST-1, either spatially or spectrally, subpixel event
repositioning allowed us to exclude approximately half the emission from
HST-1 using a region with radius $0.3\asec$ centred on the peak.  The nucleus is brighter
than HST-1 by a factor of $\sim4$.  Therefore the PSF wings from HST-1
will marginally increase the measured flux from the nucleus.  This
will contribute to the uncertainty in the subtraction of the nuclear
PSF, which is evaluated in section \ref{sec:psfsub}.  The analysis was
restricted to the angular range $60-330^{\circ}$ from W due to the
incomplete subtraction of the jet knots including HST-1.

\subsection{Nuclear PSF simulation}
\label{sec:subpsf}

The nuclear point source in M87 is brighter than the underlying
extended emission from the galaxy's hot atmosphere by more than an
order of magnitude.  The wings of the nuclear PSF contribute
significantly to the total emission to a radius of a few arcseconds.
Whilst the hot atmosphere dominates beyond this, the Bondi sphere is
located at a radius of $\sim2\asec$.  If not accurately subtracted,
high energy photons from the nucleus may distort the gas properties
measured from spectra extracted within this region.  Following
\citet{Russell15} (see also \citealt{Russell10,Siemiginowska10}), we
extracted a spectrum of the nuclear emission from a region with a
radius of $1\asec$ and determined the best-fit powerlaw model using
\textsc{xspec} (\citealt{Arnaud96}).  This model was input to the
\textit{Chandra} ray-tracer (ChaRT; \citealt{Carter03}), which traces
rays through the \textit{Chandra} mirrors.  The \textsc{marx} software
(\citealt{Davis12}) was then used to project these rays onto the
ACIS-S detector and produce a simulated events file of the nuclear
emission.  This allowed us to determine the contribution of the
nuclear point source to each spectrum extracted from different regions
of the underlying hot gas emission.  Although it was not possible to
cleanly separate the nuclear and extended emission, either spectrally
or spatially, our method removes the variable nuclear emission so that
only the non-varying cluster emission remains.  We show that the
resulting cluster surface brightness profiles extracted from
observations taken in 2000, 2010 and 2016 are consistent within the
uncertainties for different energy bands.  This method is described in
detail in sections \ref{sec:pileup}, \ref{sec:nucspec},
\ref{sec:chart} and \ref{sec:psfsub}.

\subsubsection{Pileup}
\label{sec:pileup}
\begin{figure*}
\begin{minipage}{\textwidth}
\centering
\includegraphics[width=0.4\columnwidth]{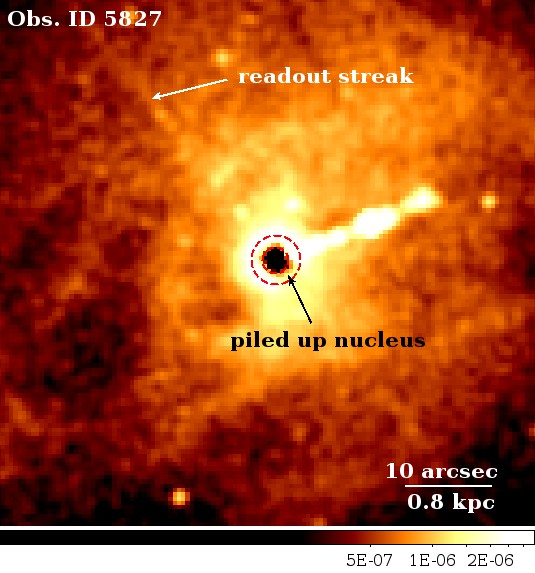}
\hspace{0.5cm}
\includegraphics[width=0.4\columnwidth]{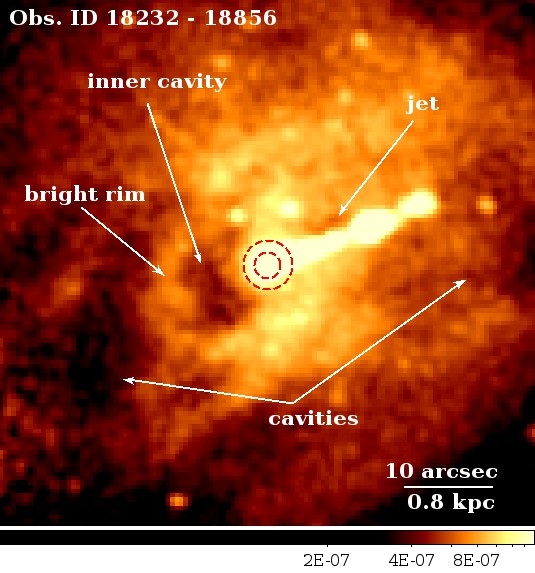}
\caption{Exposure-corrected \textit{Chandra} images for the energy range $0.5-7\keV$.  The Bondi radius at $0.12-0.22\kpc$ ($1.5-2.8\asec$) is shown by the red dashed circles.  Left: Obs. ID 5827 was taken in May 2005, close to the brightness peak of HST-1, and therefore exhibits strong pileup.  Nuclear events were so piled up that they exceeded \textit{Chandra}'s energy cutoff and were not telemetered to the ground, which produced the apparent hole.  A readout streak also runs from top left to bottom right.  Right: The new deep short frame-time \textit{Chandra} observation shows a nucleus with at most a few per cent pileup and therefore no significant distortions. The colour bar units are \expmapcorr.}
\label{fig:comparison}
\end{minipage}
\end{figure*}

The \textit{Chandra} observations of M87 are affected by varying
levels of pileup.  Pileup occurs when multiple photons from a bright
source arrive in the same detector pixel within a single integration
time ($3.1\s$ for a typical \textit{Chandra} observation).  The
photons are detected as a single event with higher energy and a
broader charge cloud distribution (see eg. \citealt{Davis01}).
Therefore, pileup hardens the source spectrum and causes grade migration,
where events are misclassified as cosmic rays.  In extreme cases, pileup
can result in lost flux when events exceed the energy filter cutoff in
the satellite and are not telemetered to the ground (Fig. \ref{fig:comparison} left).  Strong pileup of
the nucleus occurs in standard $3.1\s$ frame time observations of M87
($\sim80\%$ of events affected in obs. ID 352) and in short $0.4\s$
frame time observations from $\sim2003-2010$ when the jet knot HST-1
flared and, at peak, was more than an order of magnitude brighter than
the nucleus (\citealt{Harris09}).  Using \textsc{pimms}, the early
short frame time observation in 2000 was determined to have a pileup
fraction of only $\sim6\%$ (obs. ID 1808) whilst the monitoring
observations from 2010 to 2012 (obs. ids $11513-14974$) are closer to
$\sim10\%$.  The drop in the nuclear brightness by roughly a factor of
$2-3$ since 2012 ensures that our large observing program in 2016 has
the lowest level of pileup at only a few per cent.  The nuclear
spectrum is therefore not significantly distorted by pileup in our new
analysis. We also later confirmed this by including a statistical
treatment of pileup in the \textsc{marx} simulations and finding no
significant difference.

% Nuclear spectrum

\subsubsection{Nuclear spectrum}
\label{sec:nucspec}
A nuclear spectrum was extracted from each separate observation using
a circular region of radius $1\asec$ centred on the nuclear peak.  The
positioning was determined using sub-pixel event repositioning as
described above (section \ref{sec:dataanalysis}).  The contribution
from the galaxy's hot atmosphere was subtracted using a background
spectrum extracted in a surrounding annulus from $2-4\asec$.  However,
this assumes that the surface brightness profile for the hot gas is
flat over the radial range to the nucleus and that the nuclear PSF
contribution to this background region is negligible.  The hot gas
emission is expected to be $\sim5\%$ of the total emission within
$1\asec$ radius whilst the nuclear PSF is at most a few per cent of
the emission from $2-4\asec$ in the new datasets.  As noted in section
\ref{sec:datared}, the measured nuclear flux may also be marginally
increased by emission from HST-1.  We analysed the effect of these
systematic uncertainties in the nuclear spectral model along with
other uncertainties in the PSF subtraction in section
\ref{sec:psfsub}.

The nuclear spectra were fit separately in \textsc{xspec} version
12.9.1 (\citealt{Arnaud96}) with an absorbed power law model
\textsc{phabs(zphabs(powerlaw))} over the energy range $0.5-7\keV$.
The Galactic absorption component was fixed to the observed value of
$1.94\times10^{20}\pcmsq$ (\citealt{Kalberla05}) and the redshift of
the intrinsic absorption component was fixed to $z=0.0044$.  All other
parameters were left free and the best-fit values were determined by
\textsc{xspec}'s modified version of the C-statistic
(\citealt{Cash79,Wachter79}).  The best fit parameters for each
observation are shown in Table \ref{tab:obs}.  The pattern of nuclear
variability appears consistent with long-term monitoring studies of
X-ray variability in M87.  \citet{Harris09} characterized the nuclear
variability as `flickering', where large amplitude variations in flux
are seen on short sampling times and smaller amplitude variations are
observed on longer sampling times.  Whilst the nuclear flux varies
significantly between the separate observations, the spectral index
and intrinsic absorption are consistent within the uncertainties.
This is consistent with the variability found for other low luminosity
AGN at the centre of clusters (eg. \citealt{Russell13}).  The nuclear
spectra from the large program datasets could therefore be fitted
together in \textsc{xspec} and this best-fit model was used as the
basis of the ChaRT simulation.

\subsubsection{ChaRT and \textsc{marx}}
\label{sec:chart}
\begin{figure*}
\begin{minipage}{\textwidth}
\centering
\includegraphics[width=0.4\columnwidth]{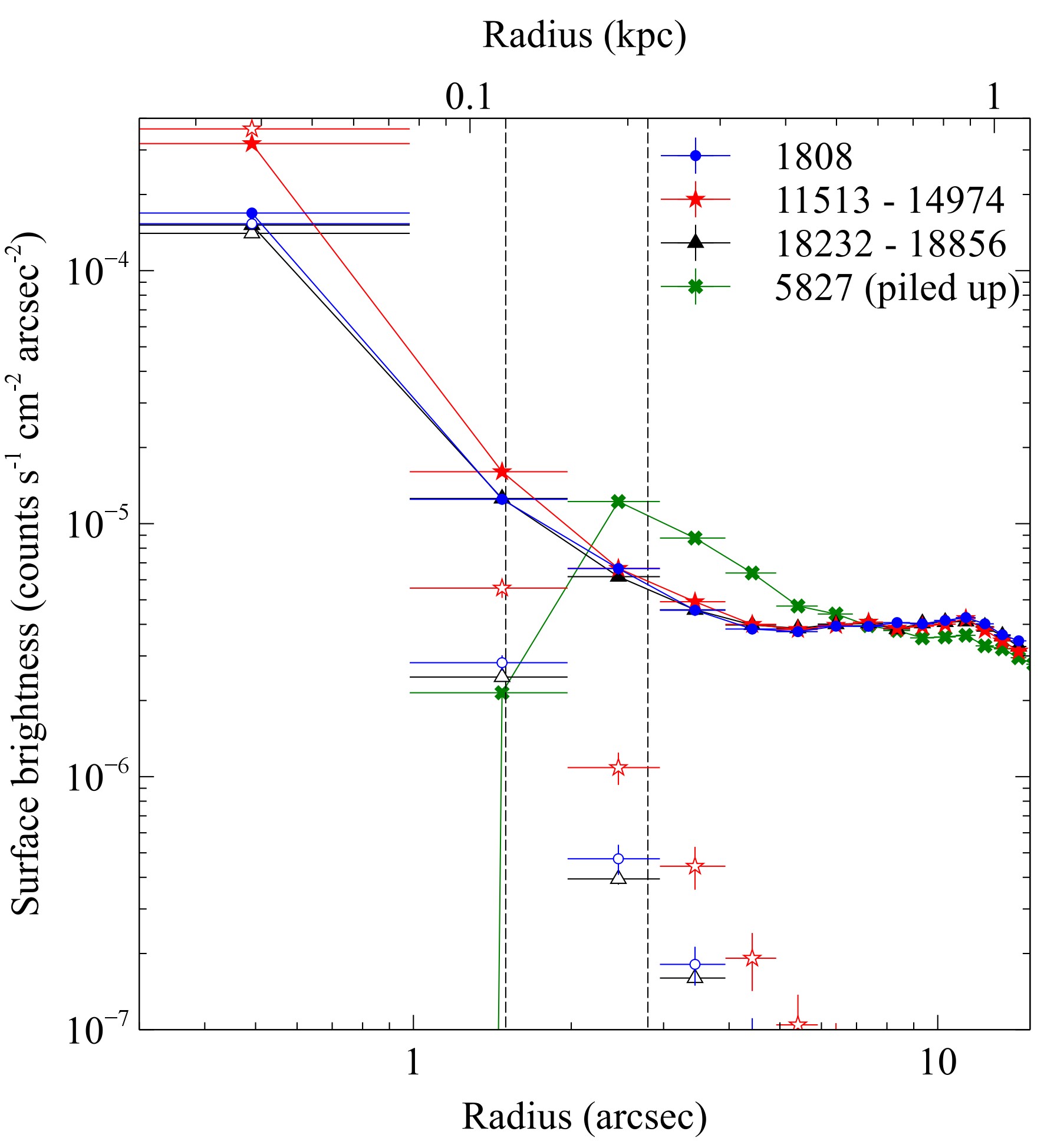}
\includegraphics[width=0.4\columnwidth]{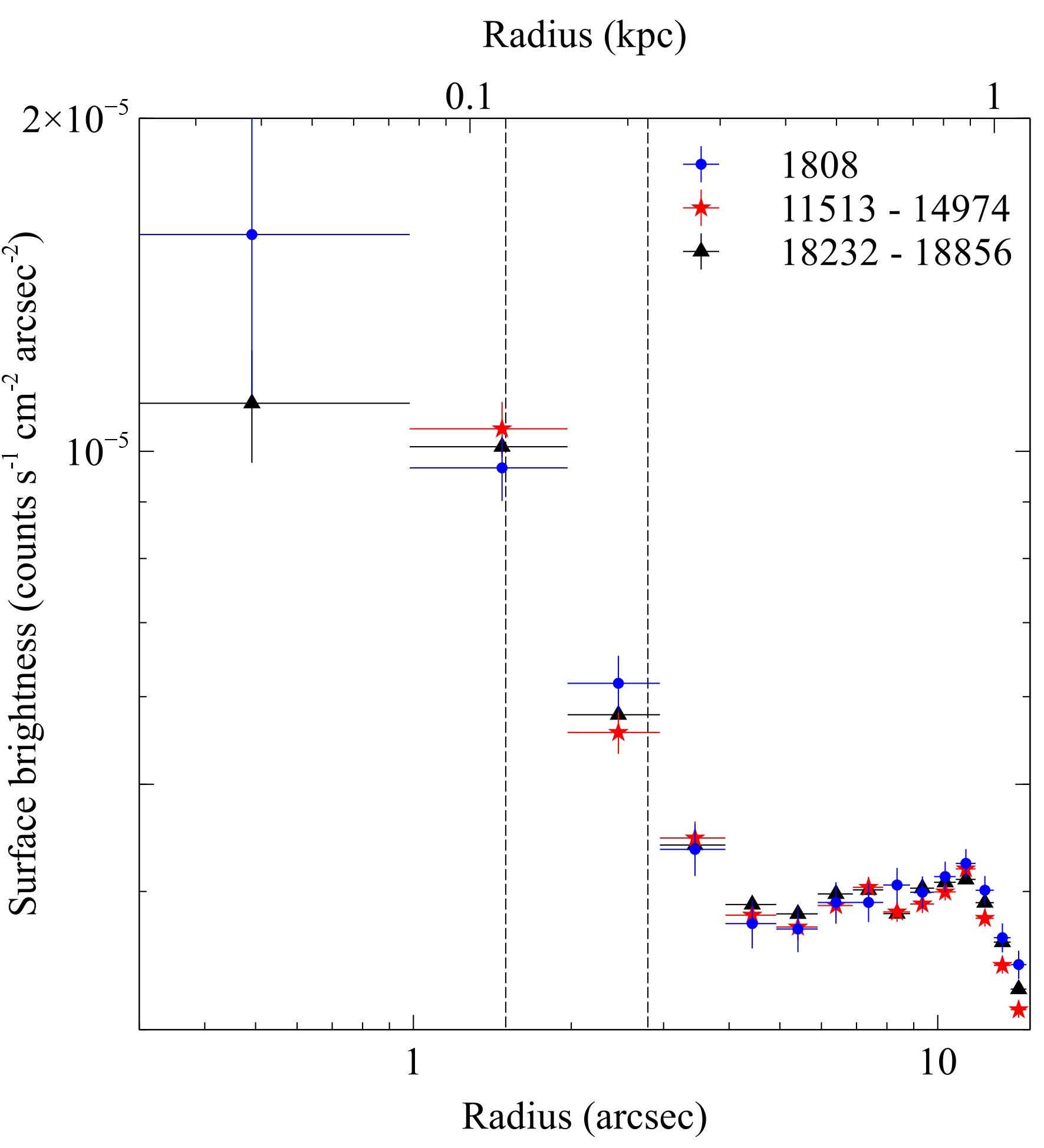}
\caption{Left: Background-subtracted surface brightness profiles in the energy range $0.5-7\keV$ for obs. ID 1808, summed obs. IDs 11513 to 14974 and summed obs. IDs 18232 to 18856 (solid points) together with the corresponding ChaRT/\textsc{marx} simulation of the nuclear PSF (open points).  Obs. ID 5827 is also shown for comparison.  This observation was taken in 2005 with a regular $3.1\s$ frame time and therefore the nucleus is strongly piled up.  Events at the centre exceeded the energy filter cutoff and were not telemetered to the ground from the satellite, which produces the sudden drop in surface brightness within $2\asec$.  Right: Surface brightness profiles for the extended hot gas emission where the nuclear PSF has been subtracted from each observation.  The radial range for the classical Bondi radius lies between the vertical dashed lines.  The profiles were extracted in a sector from $90$ to $330^{\circ}$, which excluded the jet knot emission.  The increase in surface brightness at a radius of $10\asec$ is instead due to the inner cavity's rim.  Note that vertical error bars are plotted for all points but in some cases are too small to see.}
\label{fig:subcompfull}
\end{minipage}
\end{figure*}

The ChaRT interface to the SAOTrace code was used to trace rays
through the \textit{Chandra} mirrors and \textsc{marx} then projected
these rays onto the detector.  SAOTrace was developed as a key
calibration tool by the \textit{Chandra} X-ray Center and provides the
most accurate PSF at any off-axis angle and at any energy
(eg. \citealt{Jerius95,Jerius00,Jerius02}).  Multiple ray-traces were
used to produce a total simulation exposure that was an order of
magnitude deeper than the observation and thereby ensure good photon
statistics.  \textsc{marx} version 5.3.2 was used to combine these separate
ray-traces, project them onto the ACIS-S detector and apply the
appropriate instrumental response (\citealt{Davis12}).  Following
\citet{Russell15}, we also used \textsc{marx} to determine a
correction factor for the total flux of the PSF simulation that accounts
for the fraction of the PSF lying beyond $1\asec$ radius and, for
earlier datasets, the low level of pileup.  The output
simulated events file from \textsc{marx} was used to produce simulated images and spectra
of the nuclear PSF and thereby subtract the nuclear contribution from
radial profiles of the extended gas properties.

\subsection{Subtraction of the nuclear PSF}
\label{sec:psfsub}

The nuclear and jet knot brightness vary significantly over the 16
year span of the M87 observations in Table \ref{tab:obs}.  However, if our
method subtracts this emission accurately, the surface brightness of
the remaining extended hot gas emission should be consistent within the
uncertainties.  Exposure-corrected images were produced from the
simulated events file for the energy bands $0.5-7\keV$, $0.5-2\keV$
and $2-7\keV$ with spectral weighting given by the best-fit nuclear
spectral model (section \ref{sec:nucspec}).  Surface brightness
profiles for the simulated PSF were extracted from these images using
concentric annuli with $1\asec$ width.  These simulated surface
brightness profiles were compared with, and subtracted from, observed
surface brightness profiles generated from a merged,
exposure-corrected image (obs. IDs $18232-18856$).  The X-ray
background emission was also subtracted from the observed datasets
using blank sky background images for the corresponding energy range
(section \ref{sec:dataanalysis}).  This analysis was repeated for the
stacked monitoring observations (obs. IDs $11513-14974$) and the
earliest short frame-time dataset (obs. ID 1808).  Each had a
corresponding nuclear PSF simulation.  Fig. \ref{fig:subcompfull}
compares observed, simulated and PSF-subtracted surface brightness
profiles for three epochs.  In summary, the
PSF-subtracted profiles are consistent within the uncertainties for
radii $>1\asec$.  Therefore, this method produces a reliable PSF
subtraction over this radial range.

\begin{figure*}
\begin{minipage}{\textwidth}
\centering
\includegraphics[width=0.4\columnwidth]{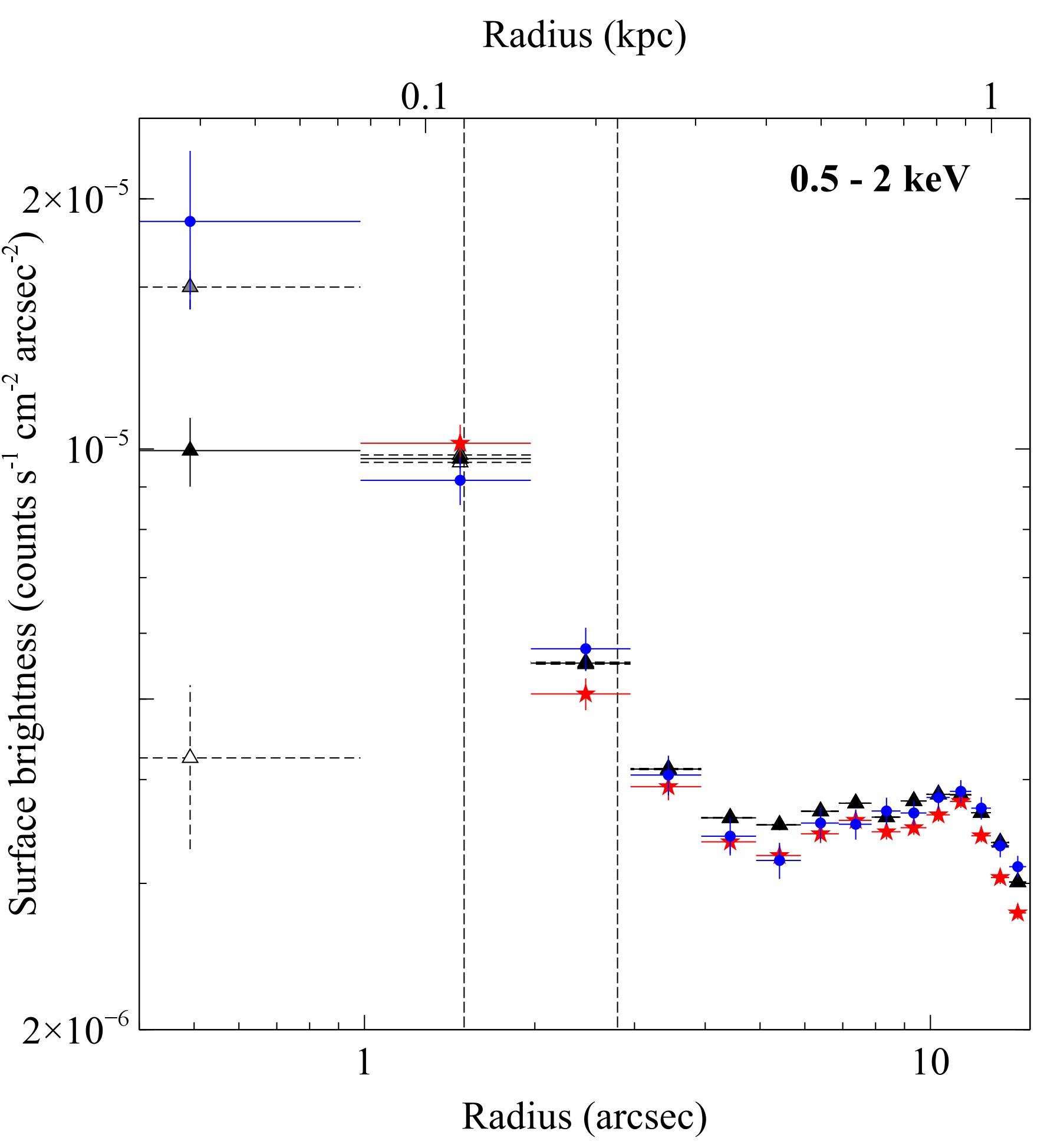}
\includegraphics[width=0.4\columnwidth]{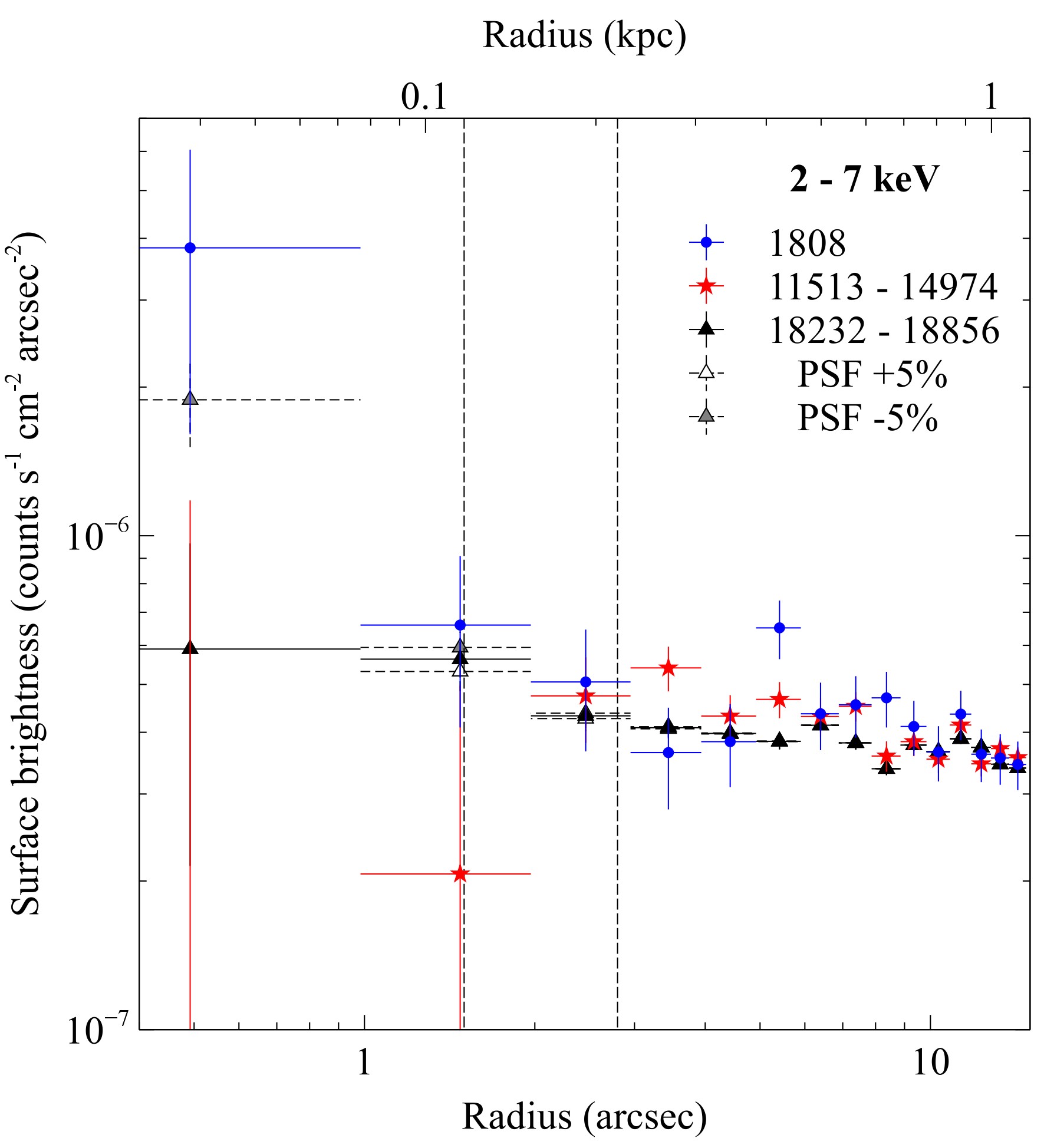}
\caption{PSF-subtracted surface brightness profiles in two energy bands, $0.5-2\keV$ and $2-7\keV$, for the obs. IDs specified in Fig. \ref{fig:subcompfull}.  The impact of under or oversubtracting the PSF simulation by 5\% is shown for obs. IDs 18232-18856 by the grey and open points, respectively.  The radial range for the classical Bondi radius lies between the vertical dashed lines.}
\label{fig:subcompE}
\end{minipage}
\end{figure*}

Fig. \ref{fig:subcompE} compares the PSF-subtracted surface brightness
profiles for soft and hard energy bands.  The temperature of the hot
atmosphere at the centre of M87 is $\sim1\keV$.  This therefore
predominantly emits in the soft energy band
(\citealt{DiMatteo03,FormanM8705,Million10}).  This gas forms dense,
clumpy structures at the galaxy's centre.  The extended emission in
the hard energy band from $2-7\keV$ is dominated by projected emission
at large radii, which doesn't exhibit strong spatial variations on
$\sim\kpc$ scales at the galaxy centre.  Therefore, in the absence of
nuclear emission, we would expect the hard band surface brightness
profile to be close to flat whilst the soft band image has a steeper
gradient that traces the low temperature gas at the galaxy centre.
The nuclear PSF is a significant emission component in both energy
bands within a few arcsec radius.  However, when it is correctly subtracted,
the hard band surface brightness profile should be roughly flat.  A
steep decline in surface brightness within a few arcsec radius would
indicate a clear oversubtraction, whilst a steep increase would
indicate an undersubtraction.  Fig. \ref{fig:subcompE} shows that the
hard band surface brightness gradient is shallow with no strong
variation within a few arcsec of the nucleus.  The soft and hard band
profiles are also consistent over all three epochs, which suggests
that the method is robust.

When the total flux of the optimum PSF simulation is increased by 5 per
cent, the $2-7\keV$ flux within $1\asec$ radius then drops below zero
at $-7\pm4\times10^{-7}\surbri$.  Alternatively, if the flux of the
PSF simulation is decreased by 5 per cent, the $2-7\keV$ flux within
$1\asec$ radius is strongly peaked at $1.9\pm0.4\times10^{-6}\surbri$.
We therefore estimate uncertainties on the total flux of the nuclear
PSF simulation to be less than 5 per cent.  This uncertainty is
ultimately due to the difficulty in cleanly separating the nuclear
emission from the background extended emission.  We tested the impact
of variations in the subtracted PSF normalization by $\pm5\%$ on the
best-fit cluster model parameters.  The resulting systematic
uncertainties were found to be insignificant compared to the
statistical uncertainties.  The nuclear PSF subtraction is therefore
sufficiently accurate for this analysis.

\begin{figure}
\centering
\includegraphics[width=0.8\columnwidth]{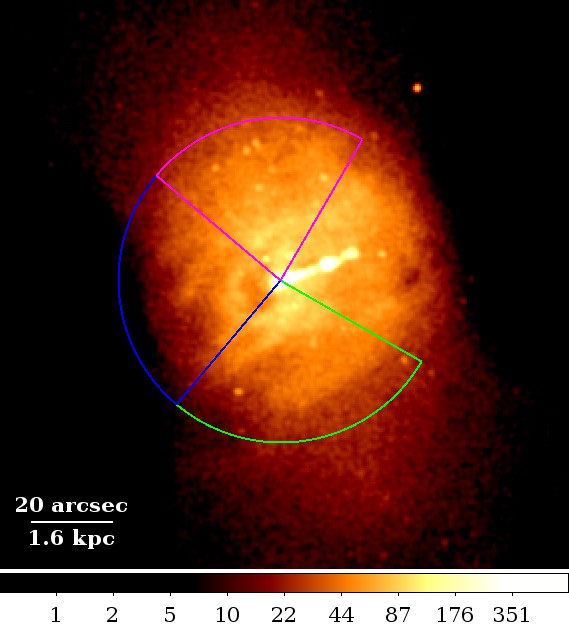}
\caption{Summed image for obs. IDs 18232 to 18856 showing the N, E and S sectors used.  The colour bar has units of counts.}
\label{fig:imgsec}
\end{figure}

\begin{figure*}
\begin{minipage}{\textwidth}
\centering
\includegraphics[width=0.9\columnwidth]{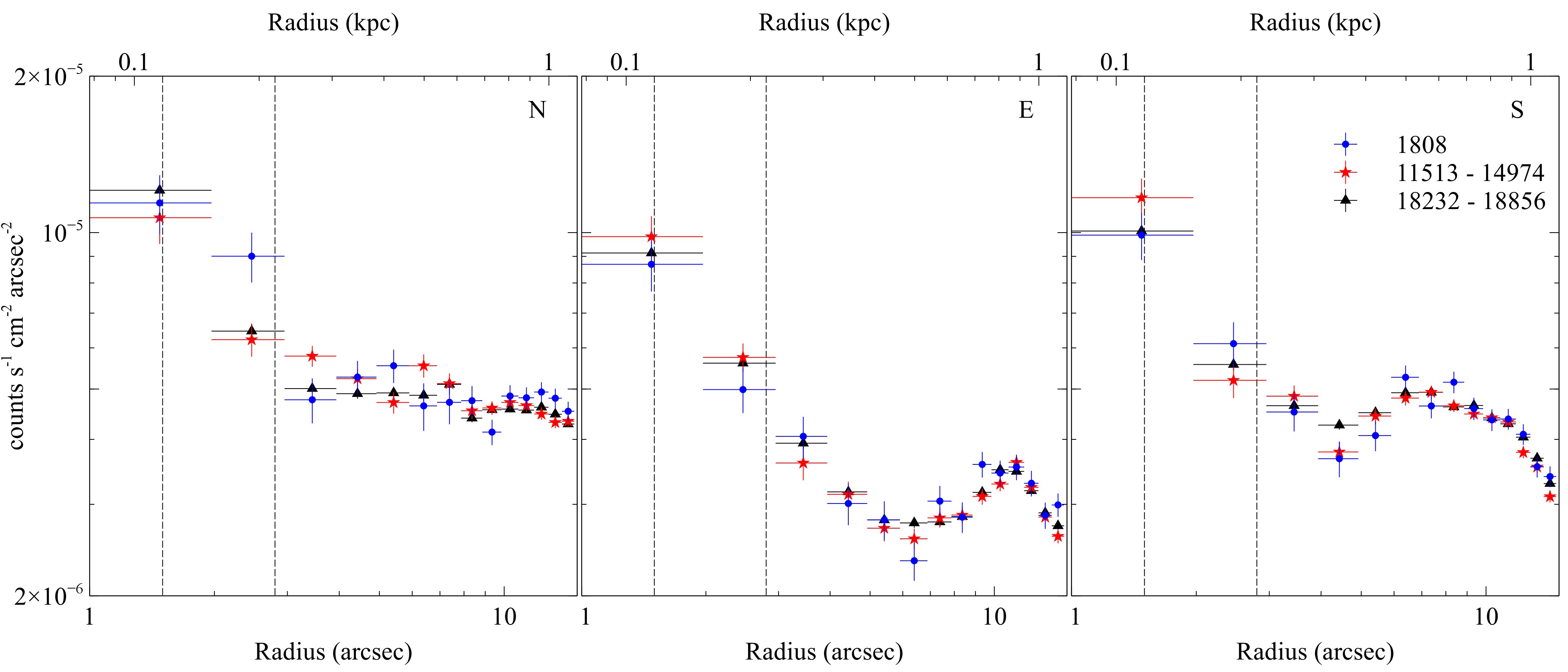}
\caption{PSF-subtracted surface brightness profiles in three sectors (Fig. \ref{fig:imgsec}) for the obs. IDs specified in Fig. \ref{fig:subcompfull}.  The radial range for the classical Bondi radius lies between the vertical dashed lines.  The drop in surface brightness due to the inner cavity can be clearly seen from $\sim4-10\asec$ in the E and S sectors.}
\label{fig:subcompsec}
\end{minipage}
\end{figure*}

The gas properties of the galaxy's hot atmosphere were also extracted
in three sectors that together covered the angular range
$90-330^{\circ}$ avoiding the jet (Fig. \ref{fig:imgsec}).
Within this range, the sectors were selected on the morphology of the
extended emission, particularly the cavity to the E of the nucleus
from $\sim4$ to $10\asec$ radius and the soft X-ray filaments.
Fig. \ref{fig:subcompsec} shows that the PSF-subtracted surface
brightness profiles are also consistent across the three epochs in
each of the sectors.  The PSF simulation was therefore accurately
centred on the observed PSF during the subtraction by the subpixel
event repositioning (section \ref{sec:dataanalysis}).  Although not
shown, the PSF-subtracted surface brightness profiles in each sector
are also consistent when split into soft and hard energy bands as
discussed above.

\subsection{Spectral analysis}
\label{sec:spec}

% similar area of detector and time therefore can merge (consistent when separately fit?)

Spectra were extracted from each observation (obs. IDs $18232-18856$) in
concentric annuli from $60-330^{\circ}$ and in sectors from
$60-140^{\circ}$ (N sector), $140-230^{\circ}$ (E sector) and
$230-330^{\circ}$ (S sector), where angles are measured anti-clockwise
from W (Fig. \ref{fig:imgsec}).  Background spectra were extracted
from the blank sky backgrounds that had been generated for each
dataset (section \ref{sec:dataanalysis}).  Observations $18232-18856$
were taken over a period of only a few months and the spatial regions
analysed are all within the central $30\asec$ and therefore have
similar responses.  Spectra from the same spatial region were
therefore summed together and the response files were averaged
according to a weighting determined from the fraction of the total
counts in each observation.  The use of summed spectra was also
verified by comparing the best-fit parameters with those determined
when fitting the individual spectra simultaneously in \textsc{xspec}.
The best-fit results in each case were found to be consistent within
the uncertainties.

The short frame-time observations were taken using a 1/8th subarray to
preserve observing efficiency.  Therefore the field of view was
limited to $1\times8\amin$.  We increased the radial range for the
analysis by extracting spectra from $14-300\asec$ in obs. ID 352,
which is only strongly piled up within the central few arcsec of the
nucleus.  Spectra were deprojected with a modified version of the
model-independent deprojection routine \textsc{dsdeproj}, which
assumes spherical symmetry and employs a geometrical procedure to
subtract off the projected emission.  This version of
\textsc{dsdeproj} includes a correction for the difference in
effective area between the summed spectra and those extracted from
obs. ID 352 (\citealt{SandersFabian07,Russell08,Russell15}).  The
spectra were analysed over the energy range $0.5-7\keV$ and counts
were grouped for at least 25 per spectral
bin.  % How many counts per spectrum?  Few thousand counts per projected spectrum (some 2500 ...)

The deprojected spectra were fitted in \textsc{xspec} version 12.9.1
(\citealt{Arnaud96}) with absorbed one, two and three component
\textsc{vapec} models (\citealt{Smith01}).  The redshift was fixed to
$z=0.0044$ and the Galactic absorption was fixed to
$n_{\mathrm{H}}=1.94\times10^{20}\psqcm$.  The $\chi^2$ statistic was
used to determine the best-fit model.  The metal abundances were fixed
relative to Fe according to the radial profiles for Fe, Si, S, Ar, Ca,
Ne, Mg and Ni measured by \citet{Million11} from the archival
\textit{Chandra} observations.  The C, N and O abundances were also
fixed relative to Fe according to the XMM-Newton RGS results of
\citet{Werner06}.  

The XMM RGS results were extracted from a $1.1\amin$ wide strip and
the \textit{Chandra} analysis excluded the inner core and the X-ray
bright arms therefore it was not clear whether the measured abundances
were applicable for the multi-phase gas within a radius of $1\kpc$.
However, in the existing data, it was not possible to constrain
individual metal abundances in the cooler gas component, even when
fixing the abundances of the spatially coincident hotter component.
We have therefore assumed no strong variation in the
metal abundances relative to iron between the multiple temperature
components, which is broadly supported by these earlier analyses.  The
abundances were measured throughout relative to the protosolar
measurements from \citet{Lodders03}.

% chi-squared and correction for PSF

An additional absorbed power-law component was used to account for the
nuclear emission in spectra extracted within a radius of $6\asec$
($0.5\kpc$).  The parameters were fixed to values determined from the
PSF simulations.  Spectra were extracted from the simulations using
identical regions, deprojected and then fit in \textsc{xspec} with an
absorbed power-law model \textsc{phabs(zphabs(powerlaw))}.  The
Galactic column density and redshift were fixed but all other
parameters were left free when fitting to the simulated spectra.  The
best-fit parameters from the simulation were then fixed in the nuclear
emission component used to model the observed spectra.  These
parameters must be fixed when modelling the observed spectra to ensure
that the PSF contribution is not incorrectly interpreted as a hot thermal component when
a multi-temperature model is used.  

The temperature and normalization of each thermal \textsc{vapec}
component were left free and the iron abundance was left free for
spectra extracted from larger radial bins.  For the multi-temperature
models, the iron abundance was tied between the different components.
The densities were calculated from the \textsc{vapec} normalizations and assumed
pressure equilibrium between the multi-temperature components.

%% F-test used to determine if extra components were necessary

\section{Results}
\label{sec:res}

\begin{figure}
\centering
\includegraphics[width=0.9\columnwidth]{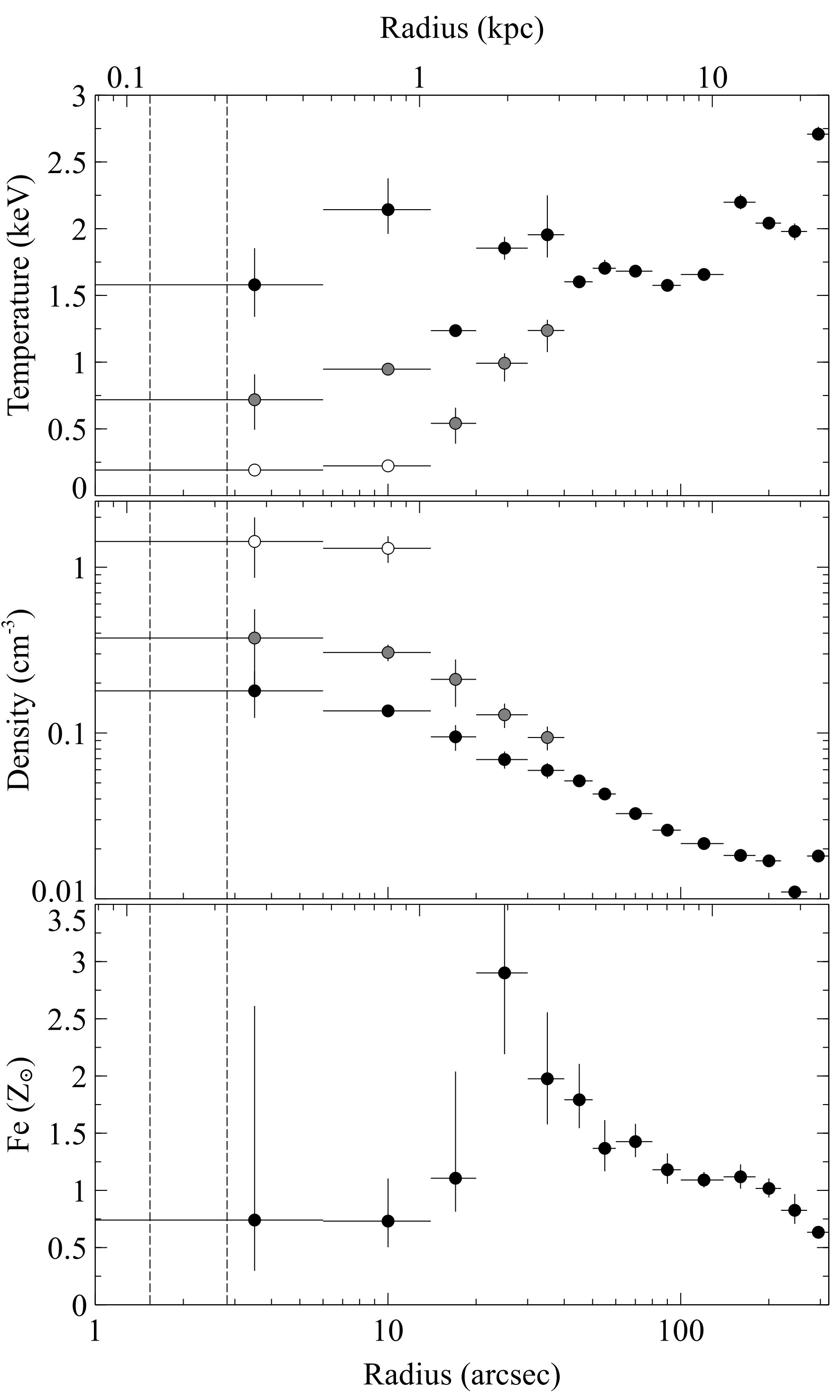}
\caption{Deprojected temperature, electron density and Fe abundance profiles for full annuli in broad radial bins.  Additional spectral components were added where these significantly improved the best-fit result and these are shown as grey and white points.  The radial range for the classical Bondi radius lies between the vertical dashed lines.}
\label{fig:fitforFe}
\end{figure}

By simulating and modelling the spectral contribution of the bright
nucleus, the temperature and density structure of the galaxy's hot
atmosphere could be extracted in sectors to a radius of $1\asec$
($78\pc$), which is within the SMBH's gravitational sphere of
influence.  Fig. \ref{fig:fitforFe} shows an analysis in large radial
bins for annuli from $60-330^{\circ}$ where up to three temperature
components were included.  The iron abundance was left as a free
parameter.  The hot gas is clearly multi-phase at the galaxy centre
with components at $1.6^{+0.3}_{-0.2}\keV$, $0.7\pm0.2\keV$ and
$0.19\pm0.03\keV$. The temperature of all three components decreases,
or is approximately constant, toward the nucleus.  The electron
density increases steadily with decreasing radius and the density of
the cooler component peaks above $1\pcmcu$.  Note that the increase in
density in the outermost radial bin is an artefact of deprojection.

The iron abundance peaks at $\sim2\Zsun$ at a radius of $2\kpc$
($25\asec$) and then decreases towards the galaxy centre to
$\sim0.7\Zsun$.  This abundance drop was previously found in
XMM-Newton observations (\citealt{Bohringer01}) and does not appear to
be due to the Fe bias that typically occurs when a single temperature
model is fitted to the spectrum of a multi-temperature medium
(eg. \citealt{Buote00}) or to resonant scattering
(\citealt{Mathews01,SandersFabian06}).  The metallicity was fixed
according to this profile in our subsequent analysis.

%\begin{figure*}
%\begin{minipage}{\textwidth}
%\centering
%\includegraphics[width=0.95\columnwidth]{sec4and8asec_fixFe.jpg}
%\caption{Deprojected temperature and electron density profiles in the N, E and S sectors (Fig. \ref{fig:imgsec}).  Additional spectral components were added where this significantly improved the best-fit result (open and semi-filled points).  The Fe abundance was fixed in each region according to the Fe abundance profile determined in full annuli (Fig. \ref{fig:fitforFe}).  The radial range for the classical Bondi radius lies between the vertical dashed lines.}
%\label{fig:secspeclg}
%\end{minipage}
%\end{figure*}

%Temperature keeps going down in all sectors, including E sector along the jet axis (Fig. \ref{fig:fitforFe} and Fig. \ref{fig:secspec}).  An additional low temperature component at $0.2\keV$ is significantly detected in several regions within a radius of $1\kpc$ but it is not clear to me if this is reliable (Fig. \ref{fig:innerspec}).  Density keeps going up.  N sector that has no cavity structure has particularly steep density gradient with $\rho{\propto}r^{-1.7}$.  Temperature structure is similar in all three sectors within the Bondi radius.  Metallicity decrease in the central kpc when fit with three temperature \textsc{vapec} model.

% UV excess in M87 eg. Buote 2001 (0.2keV = 2.3E6K).  Generally referred to on large scales (possible WHIM origin).  
% Although EUVE and ROSAT have strong detections for M87, XMM-Newton did not confirm (Kaastra et al. 2003, Matsushita et al. 2002, see Durret et al. 2008 review)

\subsection{Multi-phase gas structure within the Bondi radius}
\label{sec:multiTsec}

These annuli with broad radial binning were sub-divided into
$1-2\asec$ widths in N, E and S sectors to resolve the temperature and
density structure within the Bondi radius (Fig. \ref{fig:imgsec} and Table \ref{tab:spec}).
The N sector covers the brightest loops of soft X-ray emission that
are coincident with H$\alpha$ filaments and extend $\sim20\asec$ north
of the nucleus (eg. \citealt{Sparks93,Sparks04}).  The E sector
encompasses the inner cavity and the knotty complex of soft X-ray and H$\alpha$
emission at a radius of $\sim35\asec$ that coincides with the outer
edge of the radio lobe (\citealt{Sparks93}).  The S sector covers the
SW corner of the inner cavity and contains bright soft X-ray emission
similar to the N sector but is relatively free of H$\alpha$ emission.
The SW corner of the inner cavity extends to the S of the
  nucleus.  It was therefore not possible to exclude this cavity from
  the S sector whilst also ensuring that the innermost region was at
  least $1\asec$ across to match \textit{Chandra}'s spatial
  resolution.  We therefore positioned the boundary between the E and
  S sectors to provide comparable deprojected count rates in the inner
  regions and therefore similar parameter constraints.

\begin{figure*}
\begin{minipage}{\textwidth}
\centering
\includegraphics[width=0.95\columnwidth]{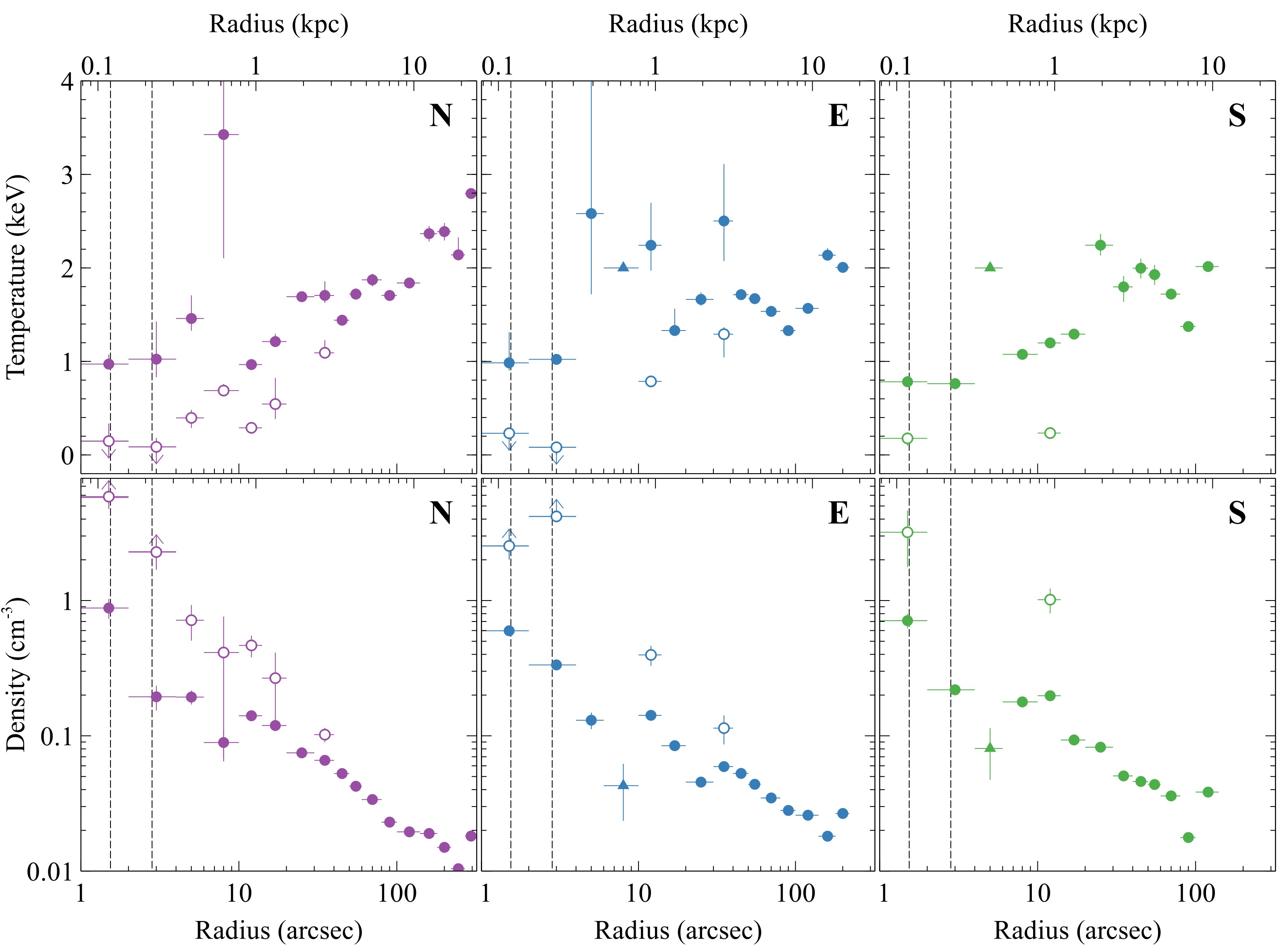}
\caption{Deprojected temperature and electron density profiles in the N, E and S sectors (Fig. \ref{fig:imgsec}).  An additional spectral component was added where this significantly improved the best-fit result (open points).  Regions affected by the cavity where the temperature was fixed are shown by the triangles.  The radial range for the classical Bondi radius lies between the vertical dashed lines.}
\label{fig:secspec}
\end{minipage}
\end{figure*}

\begin{figure*}
\begin{minipage}{\textwidth}
\centering
\includegraphics[width=0.32\columnwidth]{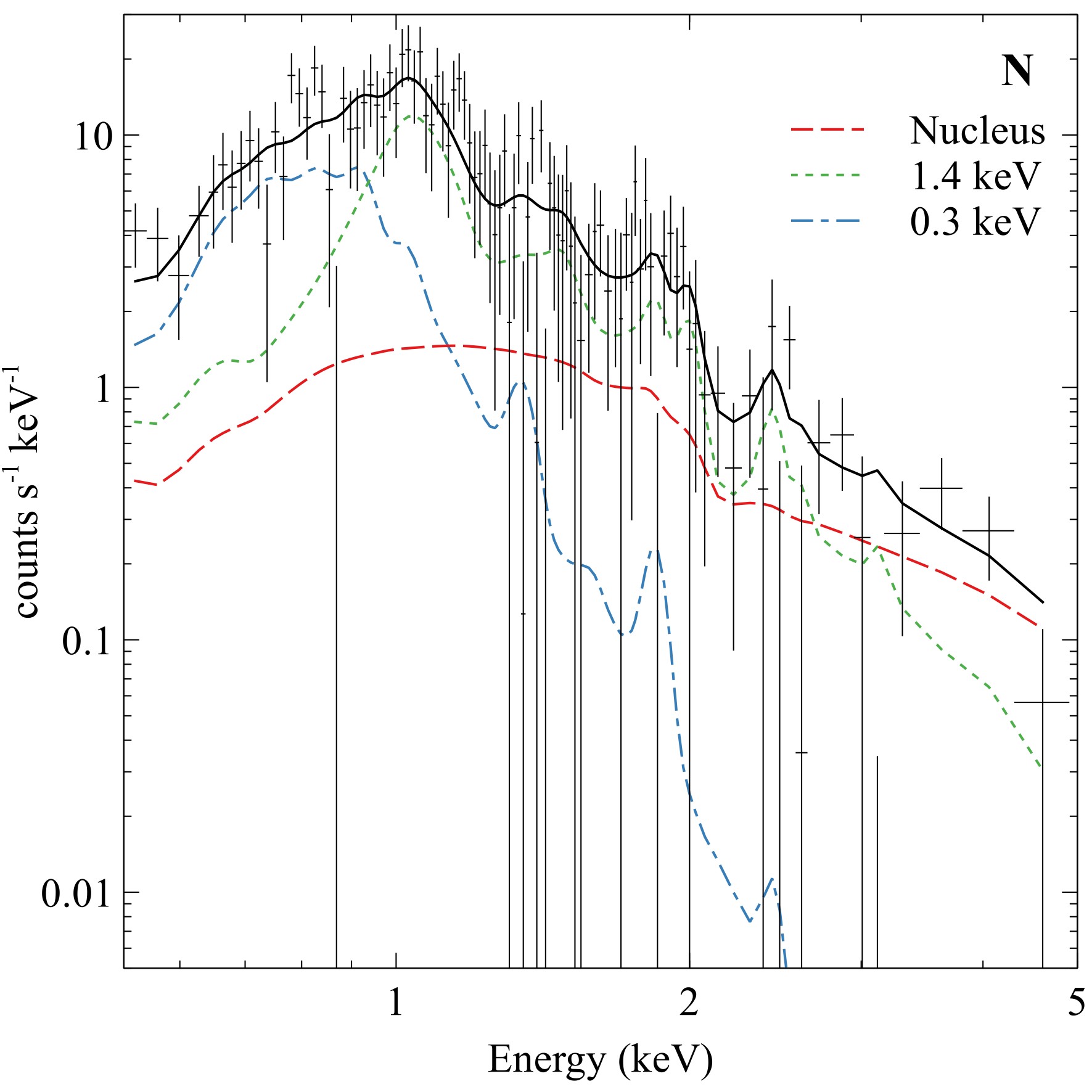}
\includegraphics[width=0.32\columnwidth]{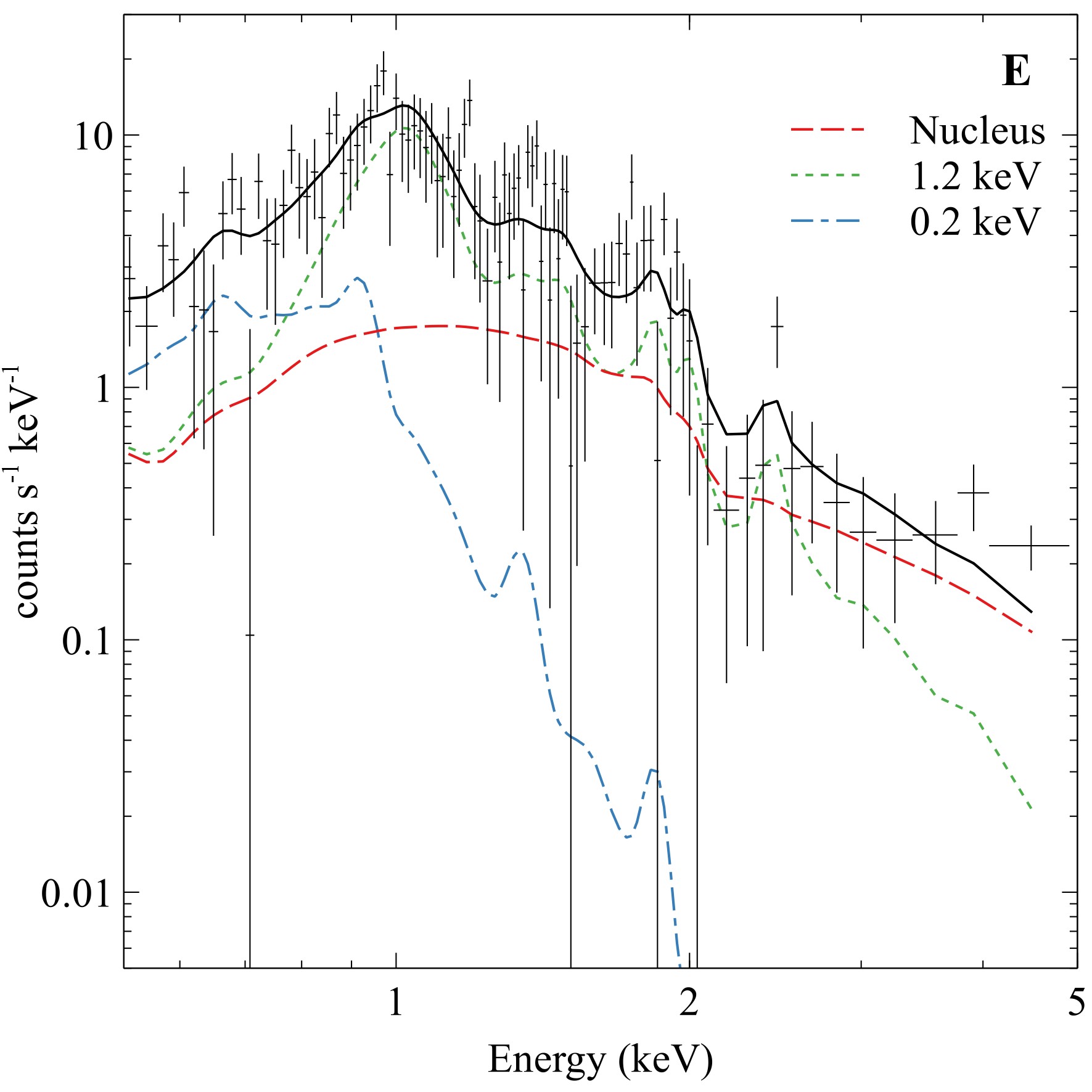}
\includegraphics[width=0.32\columnwidth]{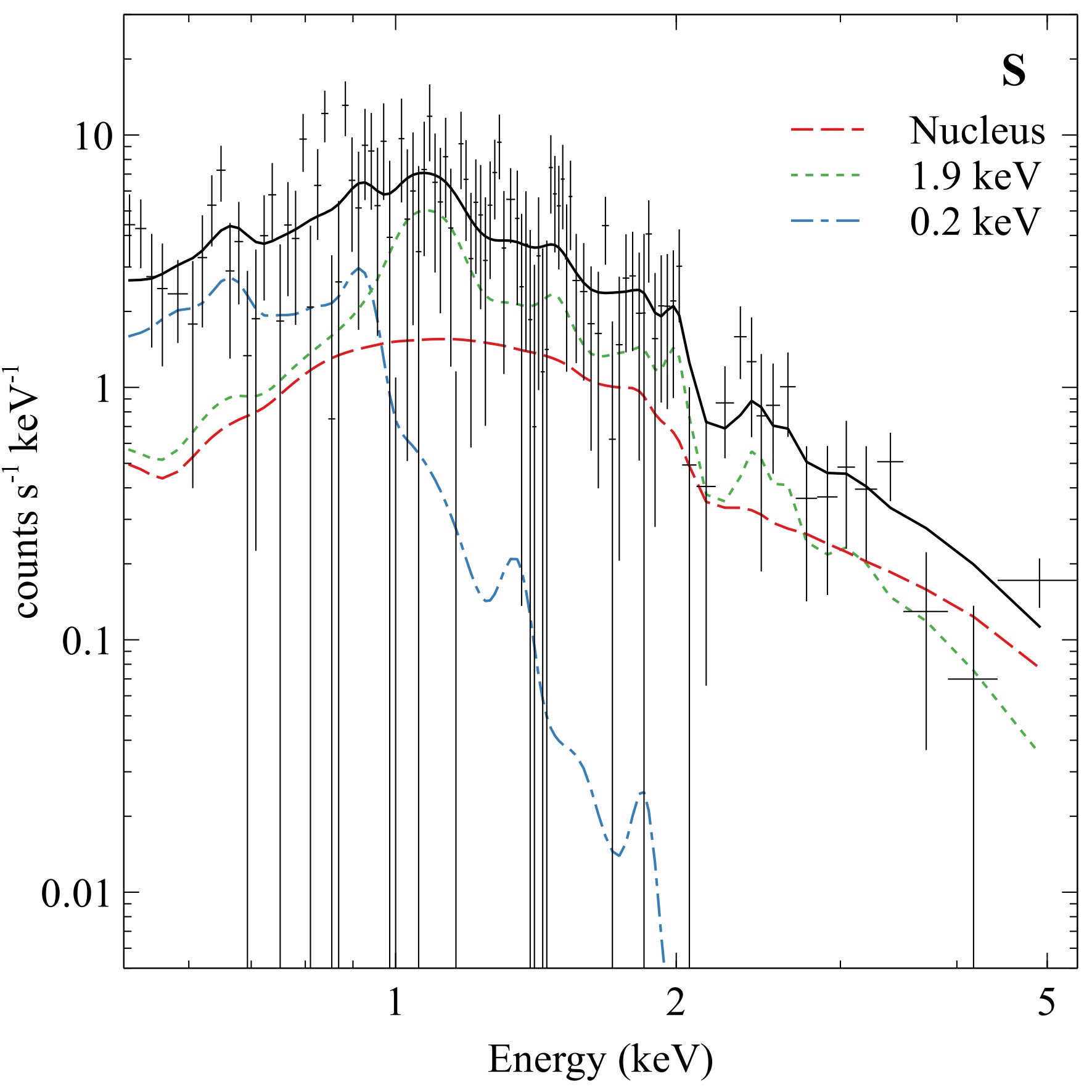}
\caption{Deprojected spectra with best-fit models for the innermost
  region ($1-6\asec$) in the N (left), E (centre) and S (right)
  sectors.  The total best-fit model is shown by the black solid line
  and each component is shown by a dashed or dotted line.  Although
  the best-fit models can be improved if some metallicity parameters
  are left free, particularly Mg and Si, the best-fit values are very
  poorly constrained.}
\label{fig:innerspec}
\end{minipage}
\end{figure*}

\begin{figure*}
\begin{minipage}{\textwidth}
\centering
\includegraphics[width=0.95\columnwidth]{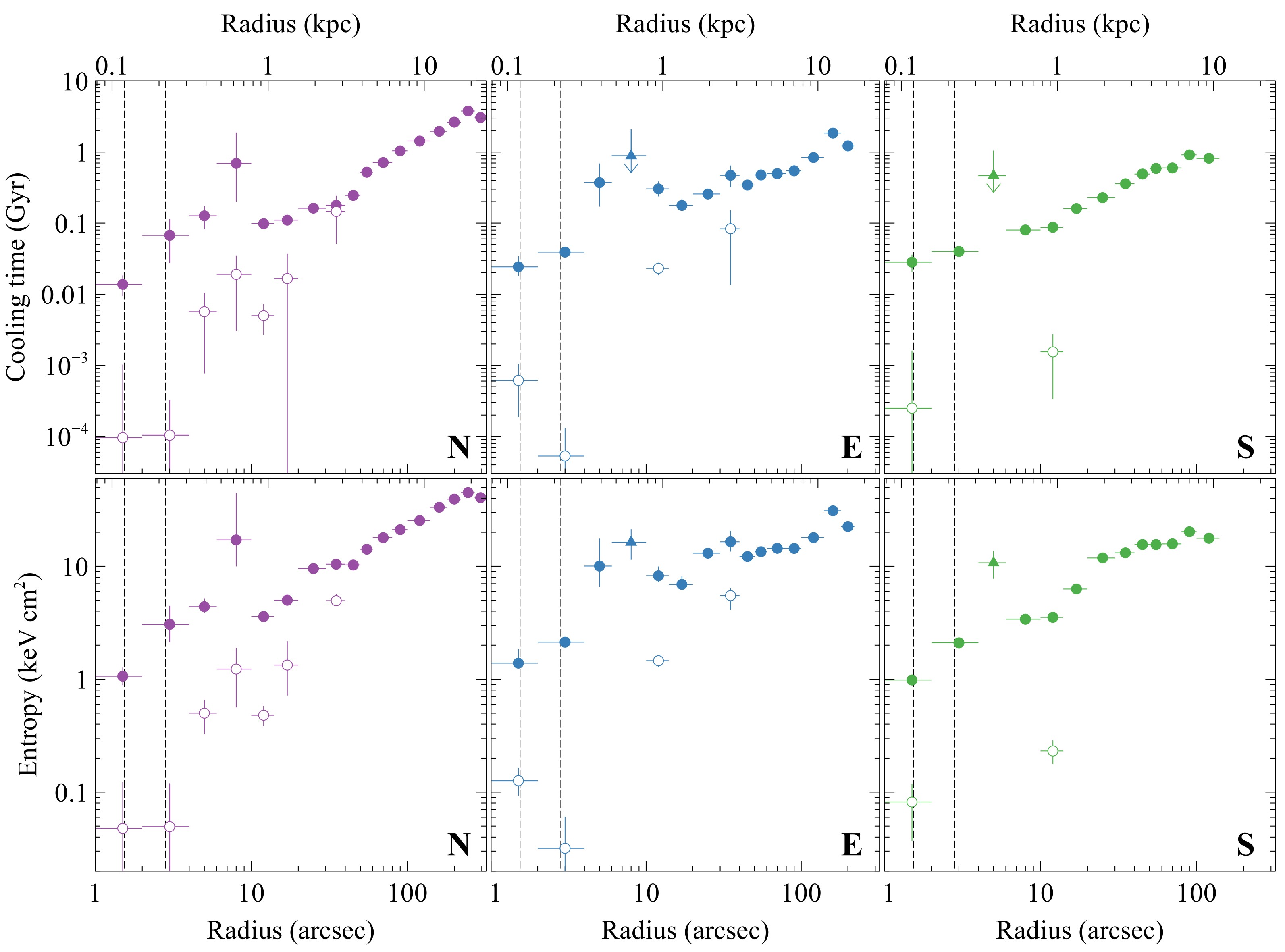}
\caption{Deprojected cooling time and entropy profiles in the N, E and S sectors (Fig. \ref{fig:imgsec}).  An additional spectral component was added where this significantly improved the best-fit result (open points).  Regions affected by the cavity where the temperature was fixed are shown by the triangles.  The radial range for the classical Bondi radius lies between the vertical dashed lines.  Within the Bondi radius, a \textsc{vmcflow} model provides a similarly good fit.}
\label{fig:sectcool}
\end{minipage}
\end{figure*}

Fig. \ref{fig:secspec} shows that the multi-temperature structure is
detected within all sectors in the central kpc with components at
$0.2\keV$ and $0.8-1\keV$.  The temperature structure within the
central few arcsec is remarkably symmetric about the nucleus, despite
the inner cavity structure.  Whilst the additional $0.2\keV$
components significantly improve the best-fit result, their specific
temperature and density values are poorly constrained by
\textit{Chandra's} $0.5-7\keV$ energy range and are therefore shown as
upper and lower limits in Fig. \ref{fig:secspec}, respectively.  No
evidence was found for a temperature increase within the Bondi radius
due to the gravitational influence of the SMBH.  Instead, the lowest
temperature gas is located closest to the nucleus.

The metallicity was fixed according to the profiles found for spectra
extracted from annuli from $60-330^{\circ}$ (section \ref{sec:res}).
We note that, if the Fe abundance is left as a free parameter for
spectra extracted in sectors, the corresponding
metallicity profiles are similar between these sectors, particularly
within the central few kpc.  However, given the large uncertainties
within $\sim10\asec$, we also investigated whether possible variations
of $\pm0.3\Zsun$ could affect the derived gas properties.  This level
of uncertainty produced negligible changes in the temperature values
for each component but shifts of $\sim15\%$ and $\sim20\%$ in the
density values for the higher and lower temperature components,
respectively.  Therefore, whilst the temperature structure is
unaffected, strong variations in the metallicity on scales of
$\sim100\pc$ could affect the measured density gradient within the
Bondi radius.

% vmcflow just as good for innermost annuli

Fig. \ref{fig:innerspec} shows the deprojected spectra and best-fit
models for regions covering $1-6\asec$ in the N, E and S sectors.  For the N
sector, the fit statistic decreases from $\chi^2=159$ with 100 degrees
of freedom for one \textsc{vapec} component to $\chi^2=98$ with 98
degrees of freedom for two \textsc{vapec} components.  The additional
low temperature component at $0.30\pm0.04\keV$ is therefore clearly
required over a single \textsc{vapec} model and fits the excess at
$\sim0.7\keV$, which is likely due to O emission
(eg. \citealt{Sakelliou02,Werner06}).  For the E sector, the fit
statistic decreases from $\chi^2=129$ with 95 degrees of
freedom for one \textsc{vapec} component to $\chi^2=109$ with 93
degrees of freedom for two \textsc{vapec} components.  Similarly, for the S sector, the fit
statistic decreases from $\chi^2=154$ with 100 degrees of
freedom for one \textsc{vapec} component to $\chi^2=103$ with 98
degrees of freedom for two \textsc{vapec} components.  The excess
emission at low energies above a $\sim1\keV$ model is less prominent
than in the N sector but still significant.  We note that the best-fit
results can be improved if some metallicity parameters are left free,
such as Mg and Si, which have prominent emission lines at $1.3$ and
$1.9\keV$ respectively.  However, the best-fit values are very poorly
constrained.

For the innermost regions within the Bondi radius, the deprojected
spectra can also be equivalently fit with an absorbed cooling flow
model \textsc{phabs(vmcflow)} with no \textsc{vapec} components
(\citealt{Mushotzky88,Johnstone92}).  For the N sector, the best-fit
cooling flow model has $\chi^2=57.2$ for 44 degrees of freedom
compared to the two \textsc{vapec} component model with $\chi^2=55.2$
for 43 degrees of freedom.  The abundances for both models were fixed
to the values determined in section \ref{sec:res}.  The best-fit upper
temperature parameter for the \textsc{vmcflow} model was
$1.5\pm0.1\keV$.  The best-fit lower temperature parameter reached the
model's lower limit at $0.08\keV$ with an upper limit of $<0.4\keV$.
The mass deposition rate was $0.028^{+0.002}_{-0.003}\Msunpyr$.
Similar results were obtained for the innermost deprojected spectra in
the E and S sectors.  The emission within the Bondi radius is
therefore equivalently consistent with a cooling flow in all three
sectors with gas cooling down out of the X-ray band from $1-1.5\keV$
at a rate of $\sim0.03\Msunpyr$.

% Also comment on constant pressure note from Nulsen?

Detection of the low temperature X-ray gas could be affected by
variations in the Galactic absorption or the growing layer of
molecular contaminant on the ACIS optical blocking filter.  However,
it is unlikely that strong variations in either of these effects would
only occur coincident with the Bondi sphere.  \citet{Lieu96} found a
smooth distribution of H\textsc{i} with
$n_{\mathrm{H}}\sim\left(1.8-2.1\right)\times10^{20}\pcmsq$ and
limited spatial variation over the field of M87 (also
\citealt{Kalberla05}).  Similarly, whilst the molecular contaminant is
not uniformly distributed with increased depth at the edges of the
filters, strong spatial variation just within the Bondi sphere at the
centre of the chip is unlikely.  Although it is possible that the
deprojection routine could compound effective area uncertainties into
the spectra for the central regions, due to the limited 1/8th field of
view of the short frame-time observations, the majority of the
projected emission was subtracted with obs. ID 352 (section
\ref{sec:spec}), which was taken early in the Chandra mission before
significant contaminant buildup.  Given the brighter nuclear emission
and shorter exposure times, it was unfortunately not possible to
verify the detection of low temperature gas in the earlier short
frame-time observations of M87.

On larger scales, the multi-temperature X-ray structure extends to
$\sim20\asec$ radius in the N sector, which is consistent with the
extent of the ionized gas filaments (eg. \citealt{Sparks93,Sparks04}).
In the E sector, the low temperature components are detected around
the periphery of the inner cavity.  The S sector is largely free of
ionized gas and we similarly find only limited low X-ray temperature
gas beyond that detected within the Bondi radius.  The narrow radial
binning produced two regions that were strongly affected by the X-ray
surface brightness depression of the inner radio bubble.  The
corresponding deprojected spectra had very low numbers of photons
therefore the temperature was fixed in these regions to that of the projected emission
at $2\keV$ (shown as triangles in Fig. \ref{fig:secspec}).  These
radial bins are included in the figures for reference but were not
considered further in our analysis.

% Density gradient of hotter component

The low temperature component likely consists of rapidly cooling blobs
and filaments within the hotter, volume-filling component at
$1-2\keV$.  The filling fraction of the cold gas is $5-20\%$ within
the Bondi sphere and roughly a few per cent at larger radii.  By
assuming that the two gas phases are in pressure equilibrium, and
ignoring magnetic pressure, the gas density of each component can be
calculated.  The density of both temperature components continues to
rise towards the nucleus.  The density of the hotter component peaks
at $0.9\pm0.1\pcmcu$ in the N sector within the Bondi radius.  The gas
density in the other sectors is similarly high within the Bondi radius
and peaks at $0.60\pm0.05\pcmcu$ and $0.71\pm0.09\pcmcu$ in the E and
S sectors, respectively.  Although poorly constrained by
\textit{Chandra's} energy range, the cooler component likely reaches
densities at least a factor of a few greater than this at $>2\pcmcu$.

The radiative cooling time, $t_{\mathrm{cool}}=(5/2)nkT/n^2\Lambda$,
and the entropy, $S=T/n_{\mathrm{e}}^{2/3}$, of each component were
calculated from the deprojected temperature and density profiles.
Note that this definition of cooling time is $5/3$ times that used in
\citet{Russell15}.  Fig. \ref{fig:sectcool} shows the radiative
cooling time and entropy profiles in each of the three sectors.  For
the higher temperature component, the cooling time and entropy
continue to decline to the Bondi radius to $\sim0.02\Gyr$ and
$\sim1\keVcmsq$, respectively.  The cooling time of the low
temperature gas drops to only $0.1-0.5\Myr$.  This is comparable to
the free fall time at this radius, which can be determined from models
of the mass distribution in M87 that include the stellar mass, the
dark matter halo and the central supermassive black hole
(eg. \citealt{Gebhardt09}).  The entropy of this component is
equivalently below $0.1\keVcmsq$.

\subsection{Density gradient within the Bondi radius}
\label{sec:densitygrad}

\begin{figure*}
\begin{minipage}{\textwidth}
\centering
\includegraphics[width=0.95\columnwidth]{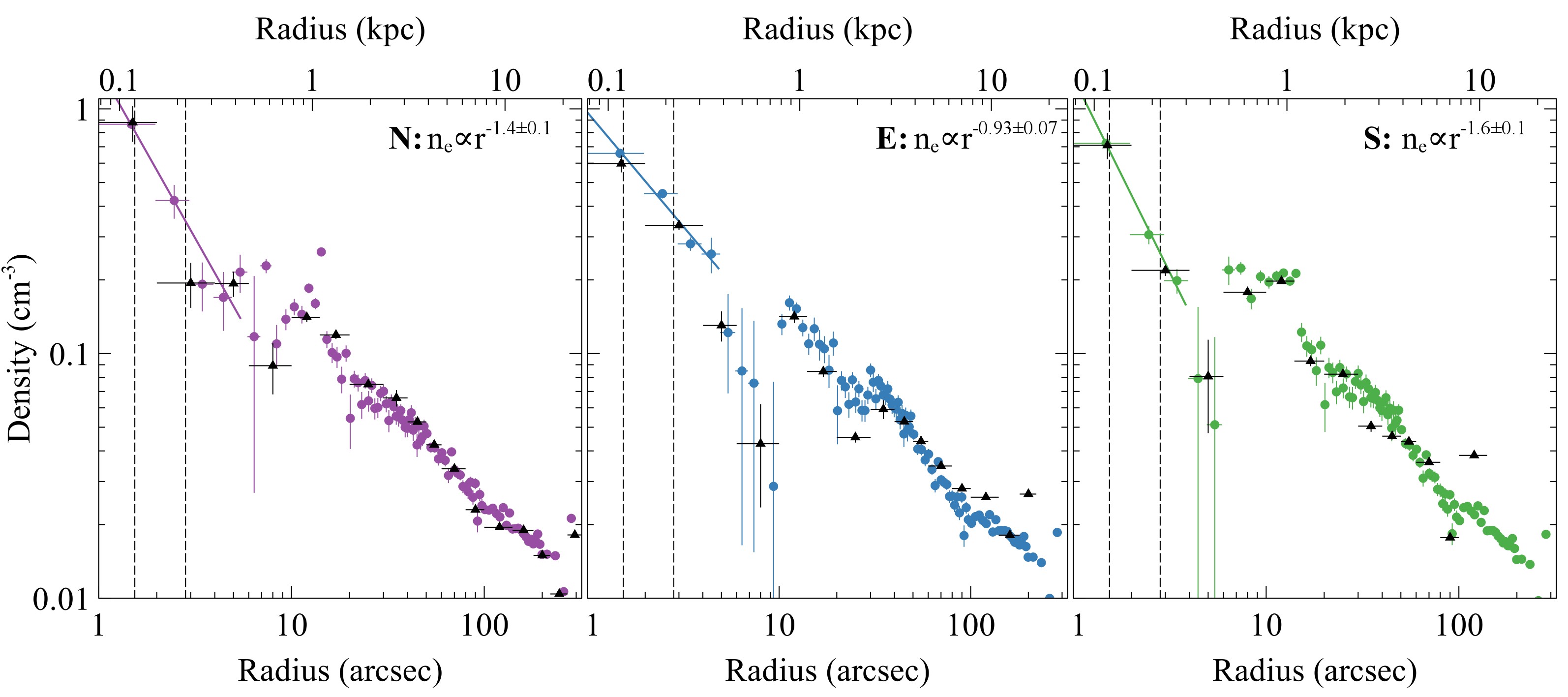}
\caption{Deprojected electron density profiles with finer radial binning and the original binning (black triangles) for the N, E and S sectors (Fig. \ref{fig:imgsec}).  The solid lines show the best-fit powerlaw models for the hot gas within the Bondi radius.  The radial range for the classical Bondi radius lies between the vertical dashed lines.}
\label{fig:densityhighres}
\end{minipage}
\end{figure*}

Density profiles with finer radial binning were also generated from
the PSF and background subtracted surface brightness profiles in each
sector.  Following eg. \citet{Cavagnolo09}, this technique
incorporates temperature and metallicity variations determined from
the spectral fitting results.  Fig. \ref{fig:densityhighres} shows the
density profiles in each sector for the hotter volume filling
component (see also Table \ref{tab:dens}).  The density values determined in narrower radial bins are
consistent with our previous results from spectral fitting.  The
density gradient across the Bondi radius was determined by fitting a
power-law model to the density profile in each sector using data
points within a radius of $4-5\asec$ ($0.3-0.4\kpc$).  This region was
selected to cover the symmetric structure about the nucleus and avoid
the inner cavity.  Note that this covered only 5 radial points in the
N sector, 4 points in the E sector and 3 points in the S sector.  The
best-fit powerlaw slopes, $n_{\mathrm{e}}{\propto}r^{\alpha}$, in the
N, E and S sectors are $\alpha=-1.4\pm0.1$, $\alpha=-0.93\pm0.07$ and
$-1.6\pm0.1$, respectively.  The E sector along the jet has a
significantly flatter density gradient than the N and S sectors that
are perpendicular to the jet.  \citet{Russell15} previously determined
a powerlaw slope of $-1.2\pm0.2$ for circular annuli, which is roughly
consistent with our new results.  The shallower observations utilized
in this earlier work prevented us from excluding the inner cavity or
separating the density profiles into separate sectors.

\subsection{Bondi accretion rate}
\label{sec:bondi}

The Bondi model states that the galaxy's hot gas atmosphere will be
accreted by the SMBH if it falls within the Bondi radius, where the
gravitational potential of the SMBH dominates over the thermal energy
of the gas (\citealt{Bondi52}).  This model assumes spherical
accretion onto a point mass, an absorbing inner boundary condition and
negligable angular momentum.  The Bondi rate can then be calculated
from the gas electron density, $n_{\mathrm{e}}$, and temperature, $T$,
and the mass of the black hole, $M_{\mathrm{BH}}$.  The Bondi
accretion rate is given by

\begin{equation}
\dot{M}_{\mathrm{B}} = \frac{4\pi\lambda\left(GM_{\mathrm{BH}}\right)^{2}}{c_{\mathrm{s}}^{3}},
\end{equation}

\noindent where $c_{\mathrm{s}}$ is the sound speed in the gas and $\lambda$ is dependent on the adiabatic index of the gas.  Assuming an adiabatic index $\gamma=5/3$, this coefficient $\lambda=0.25$.  The Bondi accretion rate can then be written as

\begin{equation}
\frac{\dot{M}_{\mathrm{B}}}{\Msunpyr}=0.012\left(\frac{k_{\mathrm{B}}T}{\mathrm{keV}}\right)^{-3/2}\left(\frac{n_{\mathrm{e}}}{\pcmcu}\right)\left(\frac{M_{\mathrm{BH}}}{10^9\Msun}\right)^2,
\end{equation}

\noindent and the Bondi radius is given by

\begin{equation}
\frac{r_{\mathrm{B}}}{\mathrm{kpc}}=0.031\left(\frac{k_{\mathrm{B}}T}{\mathrm{keV}}\right)^{-1}\left(\frac{M_{\mathrm{BH}}}{10^{9}\Msun}\right).
\end{equation}

% Black hole mass discussion

\noindent Stellar-dynamical and gas-dynamical measurements of the black hole
mass at the centre at M87 typically differ by a factor of two and are
discrepant at the $2\sigma$ level.  From stellar dynamics,
\citet{Gebhardt11} measure
$M_{\mathrm{BH}}=\left(6.6\pm0.4\right)\times10^{9}\Msun$ (see also
\citealt{Sargent78,Oldham16}).  Whilst from gas dynamics,
\citet{Walsh13} measure
$M_{\mathrm{BH}}=\left(3.5^{+0.9}_{-0.7}\right)\times10^{9}\Msun$ (see also
\citealt{Harms94,Macchetto97}).  Properties that are dependent on the
black hole mass, such as the Bondi radius and Bondi accretion rate,
are therefore shown as a range of values that span these two
measurements.

% Bondi radius for hot component and cold component!

For the volume-filling, higher temperature component, the deprojected
temperature at the centre of M87 is $0.97^{+0.10}_{-0.04}\keV$,
$0.98^{+0.33}_{-0.08}\keV$ and $0.78^{+0.06}_{-0.04}\keV$ in the N, E
and S sectors, respectively.  The Bondi radius is therefore located at
$0.12-0.21\kpc$ ($1.5-2.7\asec$) in the N and E sectors and
$0.14-0.26\kpc$ ($1.8-3.3\asec$) in the S sector.  The Bondi accretion
rate is then $0.1-0.5\Msunpyr$ for the higher temperature component.  

% We can also estimate the Bondi accretion rate for the lower
% temperature component as $0.05-0.15\Msunpyr$ (for a filling fraction
% of 10\%).  For the lower temperature component at $\sim0.2\keV$, the
% Bondi radius is at $0.5-1.4\kpc$ ($6.4-18\asec$), where the range
% reflects the uncertainty on both the black hole mass and the
% measured temperature.

% Map with free temperatures?

% Why does gas suddenly go multiphase at a particular radius??  It's essentially where the density gradient flattens.

%\subsection{Fe K$\alpha$ line emission region limit?}

\section{Discussion}

% Spherical symmetry within Bondi radius, although cooler components extend to N

Our new $300\ks$, short frame-time dataset for M87 provides a detailed
view of the hot gas properties inside the Bondi radius of the SMBH.
On these scales, the temperature structure is spherically symmetric
about the nucleus with two temperature components at $0.2\keV$ and
$0.8-1\keV$ detected in all three sectors.  The lowest temperature and
most rapidly cooling X-ray gas in M87 is therefore located within the
Bondi radius.  The radio-jet has carved out a series of cavities in the surrounding X-ray gas, each at least few kpc across, that heat the hot atmosphere on these
larger scales (eg. \citealt{FormanM8705,FormanM8707,Forman17}).  Beyond the
Bondi sphere, additional low temperature components detected in the
hot gas are predominantly located to the N of the nucleus and clearly
coincident with the soft X-ray and bright ionized gas filaments.  These filaments
extend for several kpc to the N and E of the nucleus and are entwined
around the buoyant bubbles (eg. \citealt{FormanM8705,FormanM8707}).

The density profiles in each sector plateau at a radius of
$0.5-1\kpc$, around the inner cavity structure, and then rise steeply
within the Bondi radius to $\sim1\pcmcu$.  The N and S sectors have
the steepest gradients with $\rho{\propto}r^{-1.5}$.  The E sector
along the jet axis is significantly shallower with
$\rho{\propto}r^{-0.9}$.  This structure is consistent with
anisotropic outflow along the jet.  On larger scales, the density
structure clearly varies across the different sectors due to the
surface brightness depressions and bright rims of the radio bubbles.
Additional higher density, cool gas structures associated with the
soft X-ray filaments are embedded within the hotter volume-filling gas
and detected to a radius of several kpc.  Although the flux of the low
temperature component is poorly constrained by \textit{Chandra}'s
energy range, the density is likely to be at least a factor of a few
higher at $>2\pcmcu$.  Within the Bondi radius, the gas is rapidly
cooling to low temperatures but asymmetries in the density gradient
suggest this region is a mixture of steadily inflowing material to the
N and S and outflow along the jet axis.

% The jet does not obviously appear to influence the gas structure
% within the Bondi radius and instead is inflating radio bubbles at
% larger radii that inject energy on kpc scales.  Within the Bondi
% radius, the gas appears to be cooling to low temperatures undisturbed.

\subsection{Gas cooling within the Bondi sphere}
\label{sec:discBondi}

The radiative cooling time of the $0.2\keV$ gas within the Bondi
radius is extremely short at $<1\Myr$, which is similar to the free
fall time at this radius.  In section \ref{sec:multiTsec}, we showed
that deprojected spectra extracted from within the Bondi radius can be
fit equivalently by a two temperature thermal model with components at
$0.2$ and $1\keV$ or by a single cooling flow model with gas cooling
from $1.5\keV$ down to the lower limit of $0.08\keV$.  It appears likely
that the gas is rapidly cooling down out of the X-ray band within the
Bondi radius.  The best-fit mass deposition rate within a radius of
$2\asec$ ($0.15\kpc$) is $0.028^{+0.002}_{-0.003}\Msunpyr$.  This
cooling rate is consistent with the XMM-Newton RGS limit on gas
cooling below $0.5\keV$ of $\dot{M}<0.06\Msunpyr$
(\citealt{Werner10}), which was determined in a $1.1\amin$ wide
aperture ($5.2\kpc$).

\begin{figure}
\centering
\includegraphics[width=0.9\columnwidth]{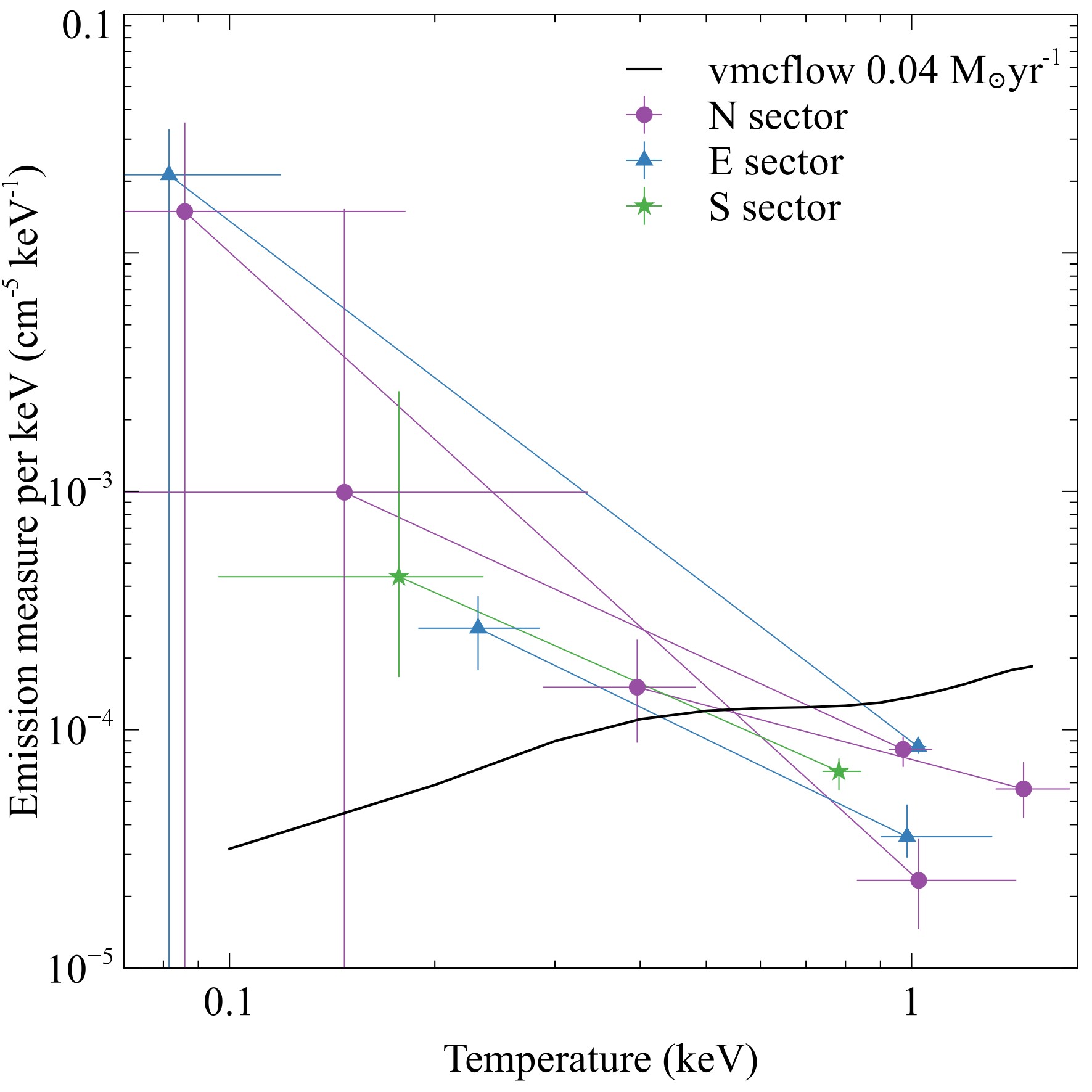}
\caption{Emission measure per unit energy as a function of temperature for the two component spectral models fit to spectra extracted from within the Bondi radius.  The black line shows the expected distribution of emission measure for a $0.04\Msunpyr$ cooling flow with the same metallicity structure.}
%\caption{Normalizations for each temperature component in a two component spectral model fit to spectra extracted from within the Bondi radius.  The black points show the expected normalizations from a two component model that was fitted to a simulated cooling flow spectrum with deposition rate $0.04\Msunpyr$.}
\label{fig:normvstemp}
\end{figure}

Fig. \ref{fig:normvstemp} shows the distribution of normalizations for
each temperature component in the two component spectral fits within
the Bondi radius compared to the expected normalization for a cooling flow model.
The gas was assumed to cool from $1.5$ to $0.08\keV$ at a rate of
$0.04\Msunpyr$ with metal abundances as detailed in section
\ref{sec:spec}.  On large scales in comparable cool core clusters,
this analysis typically demonstrates a lack of gas at the lowest
temperatures compared to the predictions of a cooling flow
(eg. \citealt{Peterson03}; \citealt{Sanders10}).  Although the normalizations for
temperature components at $<0.2\keV$ are poorly constrained by
\textit{Chandra}'s energy range, Fig. \ref{fig:normvstemp} shows that
the measured normalizations are consistent with the expectations of a
cooling flow model to low temperatures.

It is therefore plausible that the multi-phase hot atmosphere at the
centre of M87 cools catastrophically within the Bondi radius to form a
mini cooling flow.  Regardless of whether there is a continuous
distribution in temperature from $\sim1.5\keV$ down below the X-ray
band, the $0.2\keV$ gas within the Bondi radius has an extremely short
radiative cooling time that is comparable to the free fall time.  This
rapidly cooling gas will be in the form of denser, cool gas blobs that
are decoupling from any hotter volume-filling component.  It is likely
that this cooling gas has some angular momentum and would therefore
feed into the cold gas disk around the nucleus (\citealt{Ford94}).
The radius of the cold gas disk is $\sim1\asec$ ($\sim80\pc$) and
therefore lies immediately within the observed transition in the X-ray
gas properties within the Bondi radius at $1-3\asec$ ($80-250\pc$).

% Angular momentum => feeding cool disk?

HST observations of the rotating ionized gas disk found velocities
from $-500$ to $+500\kmps$ and a mass of $\sim4\pm1\times10^3\Msun$
(\citealt{Ford94,Harms94,Macchetto97}).  Whilst the bulk of the
circumnuclear gas in the disk is likely in molecular form, only
tentative detections and upper limits of typically
$\times10^{6-7}\Msun$ have so far been published
(\citealt{Braine93,Combes07,Tan08,Salome08,OcanaFlaquer10}).  ALMA
observations are expected to significantly improve on this.  Based on
the upper limits, the mini cooling flow could supply this mass of cold
gas on timescales of a few $\times10^{7-8}\yr$.  \citet{McNamara89}
determined a nuclear star formation rate within $300\pc$ radius of
$0.03\pm0.01\Msunpyr$, which suggests that a significant fraction of the
cooling gas is comsumed by star formation before it reaches the
nucleus.  % Discussion of CCA here?

% BH fuelling must be much lower than this or M87 would be a quasar?

% Source of cold clouds, feeds molecular gas??

\subsection{A settling flow within the Bondi sphere}

Assuming a steady-state, spherical flow of gas, we can use the
measurements of the gas density and temperature at the Bondi radius to
put limits on the radial flow speed and accretion rate.  In terms of
entropy, $S$, the energy equation for steady spherical flow is

\begin{equation}
\rho T v_{\mathrm{r}} \left(\frac{\mathrm{d}S}{\mathrm{d}r}\right) = n_{\mathrm{e}} n_{\mathrm{H}} \Lambda,
\label{eq:energy}
\end{equation}

\noindent where $v_r$ is the inflow speed, $\rho$ is the gas density,
$T$ is the temperature and $n_{\mathrm{e}} n_{\mathrm{H}}\Lambda$ is
the energy radiated (see eg. \citealt{Fabian77}).  Equation \ref{eq:energy} can also be written as

\begin{equation}
\label{eq:flow}
v_{\mathrm{r}} \left(\frac{\mathrm{d\ln}K}{\mathrm{d\ln}r}\right) = \frac{r}{t_{\mathrm{cool}}},
\end{equation}

\noindent where $K$ is now entropy as defined in the literature for
studies of galaxy clusters $K=k_{\mathrm{B}}T/n_{\mathrm{e}}^{2/3}$
(eg. \citealt{Voit05}).  Variations in $\mathrm{d\ln}K/\mathrm{d\ln}r$
are assumed to be negligible.

For the N sector, at a radius of $0.12\kpc$ ($1.5\asec$), the cooling
time of the $1\keV$ gas is $0.014\pm0.004\Gyr$ and the corresponding
entropy gradient $\mathrm{dln}K/\mathrm{dln}r=1.04\pm0.09$.  The
inflow velocity is then $8\pm2\kmps$.  Similarly in the S sector the
inflow velocity is $9\pm1\kmps$.  In practice, this is an upper limit
on the flow velocity because the gas may be heated and not flowing
inward.  If gas was flowing in at a higher velocity than this limit,
it would be unable to cool and the entropy gradient should be
shallower than observed.  The flow velocity within the Bondi sphere is
therefore subsonic, which indicates a settling flow rather than a
Bondi flow.  The gas flow must be supported by pressure or rotation or
both.  We therefore rule out Bondi accretion on the scale resolved at
the Bondi radius.

Mass conservation for the gas flow is given by

\begin{equation}
\label{eq:masscons}
\dot{m}_{\mathrm{acc}}=4\pi \rho v_{\mathrm{r}}r^2,
\end{equation}

\noindent where the gas density $\rho=\mu m_{\mathrm{H}}n$,
$\mu m_{\mathrm{H}}$ is the mean gas mass per particle and $n$ is the
total particle number density.  By substituting for $v_{\mathrm{r}}$
in equation \ref{eq:flow} with equation \ref{eq:masscons}, the
expected density profile can be expressed as

\begin{equation}
\rho{\propto}r^{-3/2}\left(\frac{\dot{m}_{\mathrm{acc}}T}{\Lambda\left(T\right)}\right)^{1/2}.
\end{equation}

\noindent The gas temperature profile is roughly isothermal at the
Bondi radius and should transition to $T{\propto}r^{-1}$ on smaller
scales.  For bremsstrahlung cooling, the cooling function
$\Lambda{\propto}T^{1/2}$.  Therefore, for a steady spherical flow
with constant $\dot{m}_{\mathrm{acc}}$, the density profile should
steepen from $\rho{\propto}r^{-3/2}$ at the Bondi radius to
$\rho{\propto}r^{-7/4}$ at smaller radii where the temperature
transitions to $T{\propto}1/r$.  The measured density profile slopes
to the N and S are consistent with $\rho{\propto}r^{-3/2}$ (section
\ref{sec:densitygrad}), as expected for the roughly isothermal
temperature profile.  The density gradient to the E along the jet axis
is significantly shallower at $\rho{\propto}r^{-0.93\pm0.07}$, which
likely indicates an outflow driven by the jet.  The assumption of
spherical symmetry is clearly violated by the observed density
structure within the Bondi radius.  Within the Bondi radius, the
density gradients are consistent with steady inflow to the N and S,
whilst the shallower gradient to the E indicates disruption by the jet
activity.

% Luminosity limit??

% This contradiction suggests that at least one assumption of the
% steady-state, spherical, homogeneous gas flow is violated here.  It is
% not difficult to identify possible sources for failure in these
% assumptions.  We might expect strong outflows due to the jet activity
% or winds from the accretion flow although the lack of any significant
% difference in gas properties between sectors on and off the jet axis
% suggests limited variations due to outflows.  If the rapidly cooling
% gas within the Bondi sphere is rotating, it may be dropping out of the
% hot flow and feeding the circumnuclear disk.  The gas at small radii
% may also be heated significantly.  If the flow is pressure supported,
% the temperature must rise at small radii and heat conducted outwards
% may have an appreciable effect on the gas flow at the Bondi radius
% (eg. \citealt{Shcherbakov10}).  Conductive heating from within would
% invert the entropy profile

Although equation \ref{eq:masscons} assumes spherical symmetry, which
is violated on the scale of the Bondi radius, it can provide an
estimate of the mass inflow rate.  In the N sector, the electron
density $n_{\mathrm{e}}=0.88\pm0.15\pcmcu$ at a radius of $0.12\kpc$.
Therefore, the spherical mass accretion rate
$\dot{m}_{\mathrm{acc}}\lesssim0.010\pm0.003\Msunpyr$.  The estimated
spherical mass accretion rate limit is consistent in the S sector.  We can compare
this upper limit on the accretion rate for a steady spherical inflow
with the Bondi accretion rate.  For the N sector, the Bondi rate in
the $1\keV$ gas is $0.1-0.5\Msunpyr$, where the range reflects the
uncertainty on the black hole mass (section \ref{sec:bondi}).  This is
a factor of ten higher than our upper limit on the accretion rate.  A
similar result is obtained for the E and S sectors.  This result is
consistent with the low inflow velocity, which indicates a much lower
accretion rate than the Bondi rate.

From the $PV$ work done by the radio jet to displace the hot gas in
the cavities, the mechanical jet power is
$8^{+7}_{-3}\times10^{42}\ergps$
(eg. \citealt{Reynolds96jet,Bicknell99,DiMatteo03,Rafferty06}).  The
nuclear bolometric luminosity is roughly an order of magnitude lower
at $L_{\mathrm{bol}}\sim2\times10^{41}\ergps$
(\citealt{Reynolds96,DiMatteo03}).  Assuming an efficiency of $10\%$,
our accretion rate limit of $\lesssim0.010\pm0.003\Msunpyr$ could supply
$P_{\mathrm{acc}}<6\pm2\times10^{43}\ergps$.  Therefore, the inflow
rate appears to supply roughly an order of magnitude more power than
is required for the AGN activity.

% Our results show that, at the Bondi radius, the hot gas is multiphase,
% cooling rapidly and likely transitions to a mini cooling flow.  It is
% therefore clear that a Bondi accretion rate cannot be reliably
% estimated by extrapolating the gas temperature and density from
% kpc-scales to within the Bondi sphere.

% Isn't M87 one of few objects for which Bondi accretion is sufficient?  Do we want to discuss measurement that would be made if resolution was 10 times worse?

\subsection{Outflow}
% Gradient in density, Faraday rotation measurements

Numerical simulations of hot accretion flows, such as
advection-dominated accretion flows (ADAFs,
\citealt{Ichimaru77,Rees82,Narayan94}), found that outflows are
ubiquitous and the inflow rate is expected to decrease with decreasing
radius (for a review see \citealt{Yuan14}).  This should flatten the
density profile from $\rho{\propto}r^{-1.5}$ expected for a constant
inflow rate to $\rho{\propto}r^{-p}$, where $p$ is typically $0.5-1$
(eg. \citealt{Igumenshchev99,Stone99,Stone01,Hawley02,Yuan10,Begelman12}).
The majority of the inflowing material is ejected before it reaches
the event horizon by winds off the rotationally supported accretion
flow.  Faraday rotation measurements can probe this region on scales
of tens of Schwarzchild radii, $r_{\mathrm{s}}$, and constrain the
accretion rate.  These results can be compared with X-ray
measurements, which are sensitive to the outer radial density profile.
\textit{Chandra} observations combined with Faraday rotation
measurements in Sgr A* and NGC\,3115 indicate shallow density profiles
at the Bondi radius (\citealt{Wang13,Wong14}) and reduced accretion
rates at $r\leq100r_{\mathrm{s}}$
(\citealt{Bower03,Marrone06,Marrone07,Macquart06}) with $\lesssim1\%$
of the matter captured at the Bondi radius reaching the inner regions
of the accretion flow.  In M87, the measured density profile slopes to
the N and S are consistent with $\rho{\propto}r^{-3/2}$ whilst the
gradient along the jet axis is significantly flatter with
$\rho{\propto}r^{-0.93\pm0.07}$.  This flatter density gradient
indicates an outflow directed along the jet axis.  Faraday rotation
measurements for M87 constrain the accretion rate on scales of
$\sim10r_{\mathrm{s}}$ to be an order of magnitude below our upper
limit on the inflow rate at the Bondi radius (\citealt{Kuo14}).  The
majority of the material that is gravitationally captured at the Bondi
radius is therefore lost before it reaches the SMBH, either in an
outflow or consumed by star formation (section \ref{sec:discBondi}).
We note that this analysis assumes that the observed Faraday rotation
originates in the accretion flow and that the density profile follows
a powerlaw $n{\propto}r^{-\beta}$ with $\beta\leqslant3/2$.

% In \citet{Russell15}, we found a shallow density profile at the Bondi
% radius in M87 where $\rho{\propto}r^{-1}$.  If extrapolated inwards,
% this gradient is consistent with the upper limit on the accretion rate
% at $21r_{\mathrm{s}}$ of $9.2\times10^{-4}\Msunpyr$ determined from
% Faraday rotation measurements (\citealt{Kuo14}).  The density gradient
% determined at the Bondi radius in this work is consistent with our
% earlier result, where the slopes are $-1.1\pm0.3$ and
% $-1.10^{+0.08}_{-0.09}$ in the N and E sectors, respectively (section
% \ref{sec:multiTsec}).  However, the new, deeper observations now probe
% the detailed temperature structure within the Bondi radius and show
% that the hot gas is cooling rapidly here.  Although the hot atmosphere
% may consistent of two distinct temperature components at $\sim0.2\keV$
% and $\sim1\keV$, the relative normalizations of these components
% instead suggest a transition to a mini cooling flow within the Bondi
% radius, with a continuous temperature distribution from gas cooling
% down out of the X-ray band (Fig. \ref{fig:normvstemp}).  It is
% therefore not clear whether a shallow density gradient reflects
% outflow or instead that gas is rapidly cooling out of the hot flow.

% Idea is that you trace the density gradient of the accretion inflow and then its slope tells you there must be outflow (eg. Yuan et al. 2012).  Gas may well be inflowing at Bondi radius but its not clear that we're calculating density correctly => gas is cooling out.

\subsection{Depletion of metals}

% Iron depletion due to dust?  Look at total dust mass.

The sharp drop in metal abundance within the central few kpc in M87
was first found for Fe and Si abundances in XMM-Newton observations
(\citealt{Bohringer01}).  Although the \textit{Chandra} observations
presented here cannot effectively constrain individual metal
abundances, we confirm that the metallicity drops from $2\Zsun$ to
$0.75\Zsun$ within the central $\sim1.5\kpc$.  This region is strongly
multi-phase therefore we employ up to three model temperature
components as required and find that the metallicity drop is robust.
The hot gas should be strongly enriched by centrally peaked stellar
mass loss and supernovae but instead the metallicity drops sharply.
Central declines in metallicity have been found in a number of other
cool core galaxy clusters
(eg. \citealt{Sanders02,Johnstone02,Panagoulia13,Panagoulia15}).  This
decline appears to be due to a real absence of iron rather than
resonance scattering or a spectral fitting bias.

The hot atmosphere within the central few kpc of M87 has a multiphase
structure with gas blobs cooling rapidly down out of the X-ray band
and a dusty, filamentary cool gas nebula.  The cooler phases must be
shielded from the surrounding hot X-ray gas at a few keV
(eg. \citealt{Donahue11}), presumably by magnetic fields
(eg. \citealt{Fabian08}).  Iron, and other elements such as silicon,
may adsorb onto cold dust grains (eg. \citealt{Draine09}) and
therefore are heavily depleted from the hot gas.  Measurements of the
dust mass in M87 are complicated by the strong synchrotron emission
from the jet (eg. \citealt{Perlman07,Buson09,Baes10}).  Recent
Herschel observations that cover the far IR indicate a dust mass of
$2.2\pm0.2\times10^{5}\Msun$ (\citealt{diSerego13}), which is
consistent with previous upper limits.  For \citet{Lodders03}
abundances and an average electron density of $\sim0.3\pcmcu$, we
estimate that $\sim10^{5}\Msun$ of Fe must have been depleted onto
dust grains within the central $1.3\kpc$ to explain the drop in
metallicity from $2$ to $0.75\Zsun$.  \citet{Bohringer01} found a
similar drop in the Si and O abundances.  Therefore, the total mass
that must have been depleted onto dust is very similar to the total
dust mass.

%\citet{Panagoulia13} suggested that metals locked in dust at the
%centre of the Centaurus cluster are lifted to larger radii by the
%radio bubble activity.  Dusty cool gas filaments are entrained around
%the radio lobes in M87 suggesting that they have been drawn out of the
%galaxy centre.  If the dust grains are destroyed at a radius of a few
%to ten kpc, by sputtering in the hot gas for example, this could
%explain the rapid increase in metallicity to $2\Zsun$.
%\citet{Simionescu08} showed that stellar mass loss, predominantly
%through stellar winds and type Ia supernovae, will enrich the

%% Asymmetry in metallicity decrease??  S coincident with less dust?  Not clear.
%% What is the BH fuelling mechanism?
%% Gaspari?

\section{Conclusions}

Using a new $300\ks$ short frame time \textit{Chandra} observation of
the centre of M87, we have studied the detailed structure of the hot
gas atmosphere within the Bondi radius of the supermassive black hole.
We found that:

\begin{itemize}
\item[$-$] The hot gas is multiphase on these scales with temperatures
  spanning $0.2$ to $1\keV$. The radiative cooling time of the lowest
  temperature $0.2\keV$ gas drops to only $0.1-0.5\Myr$, which is
  comparable to the free fall time.  The lowest temperature and most
  rapidly cooling gas in M87 is therefore located at the smallest
  radii that we can resolve ($\sim100\pc$).
\item[$-$] Whilst the temperature structure appears remarkably symmetric
  about the nucleus inside the Bondi radius, the density gradient is
  significantly shallower along the jet axis.  The density profiles in
  each sector analysed plateau around the inner cavity structures from
  $0.5-1\kpc$ and then smoothly steepen within the Bondi radius to
  $\sim1\pcmcu$ with cooler blobs at $>2\pcmcu$.  The best-fit
  powerlaw slopes in the N and S sectors, perpendicular to the jet
  axis, are $-1.4\pm0.1$ and $-1.6\pm0.1$, respectively, compared to
  $-0.93\pm0.07$ in the E sector along the jet axis.  The density
  structure within the Bondi radius of M87 is therefore consistent
  with steady inflow perpendicular to the jet axis and outflow
  directed along the jet axis.
\item[$-$] By putting limits on the radial flow speed
  $v_{\mathrm{r}}<8\pm2\kmps$, we rule out Bondi accretion on the
  scale resolved at the Bondi radius.  The flow velocity within the
  Bondi sphere is subsonic, which indicates a settling flow rather
  than a Bondi flow.  The gas flow must be supported by pressure or
  rotation or both.  We estimate the spherical mass inflow rate
  $\dot{m}_{\mathrm{acc}}\lesssim0.010\pm0.003\Msunpyr$.  Assuming an
  efficiency of 10\%, this inflow rate could supply roughly an order
  of magnitude more power than is required for the AGN activity.
  However, much of this material may be lost in an outflow or consumed
  by star formation before it is accreted by the SMBH.
\item[$-$] The hot gas within the Bondi radius of M87 may form a mini
  cooling flow.  The deprojected spectra extracted within the Bondi
  radius can be fit equivalently by a two temperature thermal model,
  with components at $0.2$ and $1\keV$, or by a single cooling flow
  model (no additional thermal components) with gas cooling from
  $1.5\keV$ to the lower limit of $0.08\keV$.  The best-fit cooling
  rate of $0.03\Msunpyr$ is consistent with the limit on gas cooling
  below $0.5\keV$ from the XMM-Newton RGS.  HST observations have
  shown that the circumnuclear gas disk has a radius of $80\pc$ and
  therefore lies within a putative transition to a cooling flow at
  $80-250\pc$.  If the cooling gas has significant angular momentum,
  it may feed into the circumnuclear gas disk.  Based on the upper
  limits for the cold gas mass in the disk, we estimate that the mini
  cooling flow could supply sufficient material on timescales of a few
  $\times10^{7-8}\yr$.
\item[$-$] The proposed successor missions to \textit{Chandra},
  including AXIS and Lynx, will resolve these transitions in the hot
  gas flow within the Bondi radius in many more systems and identify
  trends with jet power or molecular gas structure.  Improved
  sensitivity at low energies will also be essential to trace this
  rapidly cooling gas closer to the nucleus.
\end{itemize}

%Numerical simulations of hot accretion flows show that outflows will
%flatten the density profile from $\rho{\propto}r^{-1.5}$ expected for
%a constant inflow rate to $\rho{\propto}r^{-1}$ or shallower.

% The lower temperature gas at $0.2\keV$ is likely decoupling from the
% hot atmosphere as the radiative cooling time drops to only
% $0.1-0.5\Myr$, which is comparable to the free fall time.

\section*{Acknowledgements}

HRR and ACF acknowledge support from ERC Advanced Grant Feedback
340442.  HRR acknowledges support from an STFC Ernest Rutherford
Fellowship.  PEJN was supported by NASA contract NAS8-03060.  We thank the reviewer for their helpful and constructive comments.  This research has made use of
data from the Chandra X-ray Observatory and software provided by the
Chandra X-ray Center (CXC).  Many of the plots in this paper were made using the Veusz
software, written by Jeremy Sanders.

\begin{table*}
\begin{minipage}{\textwidth}
\caption{Temperature, electron density, Fe abundance, radiative cooling time and entropy profiles for each sector as shown in Figs. \ref{fig:secspec} and \ref{fig:sectcool}.  The metallicities were determined from spectra extracted in larger annuli and fixed to the values given here as detailed in section \ref{sec:spec}.}
\begin{center}
\begin{tabular}{l c c c c c c}
\hline
Sector & Radius & Temperature & Density & Fe & t$_{\mathrm{cool}}$ & Entropy \\
 &(kpc) & ($\keV$) & ($\pcmcu$) & ($\Zsun$) & ($\Gyr$) & ($\keVcmsq$) \\
\hline
N & $0.12\pm0.04$ & $0.15^{+0.19}_{-0.15}$ & $5.9\pm1.1$ & 0.74 & $\left(9\pm8\right)\times10^{-5}$ & $0.05^{+0.06}_{-0.05}$ \\
  &  & $0.97^{+0.10}_{-0.04}$ & $0.88\pm0.15$ & & $0.014\pm0.004$ & $1.1^{+0.2}_{-0.1}$ \\
  & $0.23\pm0.08$ & $0.09^{+0.10}_{-0.09}$ & $2.3\pm0.6$ & 0.74 & $\left(1.0^{+1.8}_{-1.0}\right)\times10^{-4}$ & $0.05^{+0.06}_{-0.05}$ \\
  &  & $1.0^{+0.4}_{-0.2}$ & $0.19\pm0.04$ & & $0.07^{+0.04}_{-0.03}$ & $3.1^{+1.3}_{-0.7}$ \\
  & $0.39\pm0.08$ & $0.40^{+0.09}_{-0.11}$ & $0.7\pm0.2$ & 0.74 & $0.006\pm0.005$ & $0.5^{+0.1}_{-0.2}$ \\
  & & $1.46^{+0.25}_{-0.13}$ & $0.19\pm0.02$ & & $0.13^{+0.05}_{-0.04}$ & $4.4^{+0.8}_{-0.5}$ \\
  & $0.62\pm0.16$ & $0.69^{+0.07}_{-0.05}$ & $0.4\pm0.3$ & 0.73 & $0.02\pm0.02$ & $1.2\pm0.7$ \\
  & & $3.4^{+5.5}_{-1.3}$ & $0.09\pm0.02$ & & $0.7^{+1.2}_{-0.5}$ & $17^{+28}_{-7}$ \\
  & $0.94\pm0.16$ & $0.29\pm0.05$ & $0.47\pm0.09$ & 0.73 & $0.005\pm0.002$ & $0.5\pm0.1$ \\
  & & $0.97^{+0.04}_{-0.05}$ & $0.141\pm0.006$ & & $0.10\pm0.01$ & $3.6\pm0.2$ \\
  & $1.33\pm0.23$ & $0.5^{+0.3}_{-0.2}$ & $0.27\pm0.15$ & 1.1 & $0.02\pm0.02$ & $1.3^{+0.8}_{-0.6}$ \\
  & & $1.21^{+0.08}_{-0.04}$ & $0.119\pm0.04$ & & $0.11\pm0.02$ & $5.0^{+0.4}_{-0.2}$ \\
  & $1.95\pm0.39$ & $1.69\pm0.03$ & $0.075\pm0.002$ & 2.9 & $0.16\pm0.01$ & $9.5\pm0.2$ \\
  & $2.73\pm0.39$ & $1.09^{+0.14}_{-0.03}$ & $0.10\pm0.01$ & 2.0 & $0.15\pm0.10$ & $5.0^{+0.7}_{-0.4}$ \\
  & & $1.71^{+0.15}_{-0.08}$ & $0.066\pm0.005$ & & $0.18^{+0.04}_{-0.03}$ & $10.5^{+1.1}_{-0.7}$ \\
  & $3.51\pm0.39$ & $1.44\pm0.04$ & $0.053\pm0.001$ & 1.8 & $0.25\pm0.02$ & $10.3\pm0.3$ \\
  & $4.29\pm0.39$ & $1.72^{+0.06}_{-0.03}$ & $0.042\pm0.001$ & 1.4 & $0.52\pm0.04$ & $14.2^{+0.5}_{-0.3}$ \\
  & $5.46\pm0.78$ & $1.87\pm0.07$ & $0.0338\pm0.0006$ & 1.4 & $0.71\pm0.04$ & $17.9\pm0.7$ \\
  & $7.02\pm0.78$ & $1.71\pm0.04$ & $0.0230\pm0.0006$ & 1.2 & $1.04\pm0.07$ & $21.1\pm0.6$ \\
  & $9.37\pm1.56$ & $1.84\pm0.06$ & $0.0195\pm0.0003$ & 1.1 & $1.43\pm0.07$ & $25.4^{+0.8}_{-0.9}$ \\
  & $12.49\pm1.56$ & $2.37\pm0.08$ & $0.0190\pm0.0002$ & 1.1 & $1.96\pm0.09$ & $33.3\pm1.2$ \\
  & $15.61\pm1.56$ & $2.39\pm0.09$ & $0.0150\pm0.0002$ & 1.0 & $2.6\pm0.1$ & $39.3^{+1.5}_{-1.6}$ \\
  & $19.12\pm1.95$ & $2.14^{+0.19}_{-0.06}$ & $0.0104\pm0.0002$ & 0.8 & $3.8^{+0.4}_{-0.2}$ & $44.9^{+4.0}_{-1.3}$ \\
  & $23.02\pm1.95$ & $2.80\pm0.05$ & $0.01814\pm0.00005$ & 0.6 & $3.06\pm0.07$ & $40.5\pm0.8$ \\ 
E & $0.12\pm0.04$ & $0.23^{+0.05}_{-0.04}$ & $2.5\pm0.5$ & 0.74 & $\left(6\pm4\right)\times10^{-4}$ & $0.12\pm0.03$ \\
  & & $0.98^{+0.33}_{-0.08}$ & $0.60\pm0.05$ & & $0.024^{+0.010}_{-0.006}$ & $1.4^{+0.5}_{-0.1}$ \\
  & $0.23\pm0.08$ & $0.08^{+0.04}_{-0.08}$ & $4.2\pm0.2$ & 0.74 & $\left(5\pm5\right)\times10^{-5}$ & $0.03^{+0.01}_{-0.03}$ \\
  & & $1.02\pm0.03$ & $0.33\pm0.01$ & & $0.039\pm0.004$ & $2.12\pm0.09$ \\
  & $0.39\pm0.08$ & $2.6^{+1.9}_{-0.9}$ & $0.13\pm0.02$ & 0.74 & $0.4^{+0.3}_{-0.2}$ & $10^{+8}_{-3}$ \\
  & $0.62\pm0.16$ & 2.0 & $0.04\pm0.02$ & 0.73 & $0.9\pm0.9$ & $16\pm5$ \\
  & $0.94\pm0.16$ & $0.79\pm0.03$ & $0.40\pm0.07$ & 0.73 & $0.023\pm0.005$ & $1.5\pm0.2$ \\
  & & $2.2^{+0.5}_{-0.3}$ & $0.142\pm0.008$ & & $0.30^{+0.08}_{-0.07}$ & $8^{+2}_{-1}$ \\
  & $1.33\pm0.23$ & $1.33^{+0.23}_{-0.04}$ & $0.084\pm0.004$ & 1.1 & $0.18^{+0.04}_{-0.03}$ & $6.9^{+1.2}_{-0.3}$ \\
  & $1.95\pm0.39$ & $1.66^{+0.08}_{-0.07}$ & $0.045\pm0.002$ & 2.9 & $0.26\pm0.03$ & $13.1\pm0.7$ \\
  & $2.73\pm0.39$ & $1.29^{+0.07}_{-0.25}$ & $0.11\pm0.03$ & 2.0 & $0.08\pm0.07$ & $5.5^{+0.9}_{-1.4}$ \\
  & & $2.5^{+0.6}_{-0.4}$ & $0.11\pm0.03$ & & $0.5\pm0.2$ & $16^{+4}_{-3}$ \\
  & $3.51\pm0.39$ & $1.71^{+0.06}_{-0.03}$ & $0.053\pm0.001$ & 1.8 & $0.34\pm0.02$ & $12.2^{+0.5}_{-0.3}$ \\
  & $4.29\pm0.39$ & $1.67^{+0.04}_{-0.05}$ & $0.044\pm0.001$ & 1.4 & $0.48\pm0.04$ & $13.5^{+0.4}_{-0.5}$ \\
  & $5.46\pm0.78$ & $1.53\pm0.03$ & $0.0347\pm0.0006$ & 1.4 & $0.50\pm0.02$ & $14.4\pm0.3$ \\
  & $7.02\pm0.78$ & $1.33\pm0.01$ & $0.0280\pm0.0004$ & 1.2 & $0.54\pm0.02$ & $14.4\pm0.2$ \\
  & $9.37\pm1.56$ & $1.57\pm0.01$ & $0.0259\pm0.0002$ & 1.1 & $0.84\pm0.02$ & $17.9\pm0.2$ \\
  & $12.49\pm1.56$ & $2.14^{+0.08}_{-0.04}$ & $0.0181\pm0.0002$ & 1.1 & $1.85^{+0.09}_{-0.07}$ & $31.0^{+1.1}_{-0.6}$ \\
  & $15.61\pm1.56$ & $2.01\pm0.01$ & $0.02665\pm0.00008$ & 1.0 & $1.22\pm0.01$ & $22.5\pm0.1$ \\
S & $0.12\pm0.04$ & $0.18^{+0.06}_{-0.08}$ & $3.2\pm1.4$ & 0.74 & $<0.001$ & $0.08\pm0.04$ \\
  & & $0.78^{+0.06}_{-0.04}$ & $0.71\pm0.09$ & & $0.028\pm0.008$ & $1.0\pm0.1$ \\
  & $0.23\pm0.08$ & $0.76\pm0.06$ & $0.22\pm0.01$ & 0.74 & $0.040\pm0.007$ & $2.1\pm0.2$ \\
  & $0.39\pm0.08$ & 2.0 & $0.08\pm0.03$ & 0.74 & $<0.6$ & $11\pm3$ \\
  & $0.62\pm0.16$ & $1.08^{+0.04}_{-0.02}$ & $0.178\pm0.003$ & 0.73 & $0.080\pm0.005$ & $3.40^{+0.13}_{-0.07}$ \\
  & $0.94\pm0.16$ & $0.23^{+0.05}_{-0.04}$ & $1.0\pm0.2$ & 0.73 & $0.002\pm0.001$ & $0.23^{+0.06}_{-0.05}$ \\
  & & $1.20\pm0.03$ & $0.198\pm0.004$ & & $0.087\pm0.006$ & $3.53\pm0.09$ \\
  & $1.33\pm0.23$ & $1.29\pm0.04$ & $0.093\pm0.004$ & 1.1 & $0.16\pm0.02$ & $6.3\pm0.3$ \\
  & $1.95\pm0.39$ & $2.2\pm0.2$ & $0.082\pm0.002$ & 2.9 & $0.23\pm0.02$ & $1..8^{+0.7}_{-0.6}$ \\
  & $2.73\pm0.39$ & $1.8^{+0.1}_{-0.2}$ & $0.051\pm0.003$ & 2.0 & $0.36^{+0.04}_{-0.06}$ & $13.2^{+0.9}_{-1.3}$ \\
  & $3.51\pm0.39$ & $2.0\pm0.1$ & $0.046\pm0.001$ & 1.8 & $0.49\pm0.05$ & $15.6\pm0.9$ \\
  & $4.29\pm0.39$ & $1.9\pm0.1$ & $0.044\pm0.001$ & 1.4 & $0.59^{+0.05}_{-0.06}$ & $15.6\pm0.9$ \\
  & $5.46\pm0.78$ & $1.72^{+0.06}_{-0.01}$ & $0.0359\pm0.0005$ & 1.4 & $0.60^{+0.03}_{-0.02}$ & $15.8^{+0.6}_{-0.1}$ \\
  & $7.02\pm0.78$ & $1.37^{+0.04}_{-0.02}$ & $0.0177\pm0.0005$ & 1.2 & $0.91^{+0.06}_{-0.08}$ & $20.2^{+0.6}_{-0.4}$ \\
  & $9.37\pm1.56$ & $2.01\pm0.01$ & $0.0384\pm0.0001$ & 1.1 & $0.816\pm0.008$ & $17.7\pm0.1$ \\
\hline
\end{tabular}
\end{center}
\label{tab:spec}
\end{minipage}
\end{table*}

\begin{table*}
\begin{minipage}{\textwidth}
\caption{Electron density profiles for each sector generated using finer radial binning as shown in Fig. \ref{fig:densityhighres}.}
\begin{center}
\begin{tabular}{l c c c}
\hline
Radius & N sector & E sector & S sector \\
 (arcsec) & n$_{\mathrm{e}}$ (cm$^{-3}$) & n$_{\mathrm{e}}$ (cm$^{-3}$) & n$_{\mathrm{e}}$ (cm$^{-3}$)\\
\hline
$0.115\pm0.038$ & $0.87\pm0.04$ & $0.66\pm0.02$ & $0.72\pm0.02$ \\
$0.192\pm0.038$ & $0.42\pm0.07$ & $0.45\pm0.02$ & $0.31\pm0.03$ \\
$0.269\pm0.038$ & $0.19\pm0.04$ & $0.28\pm0.02$ & $0.20\pm0.02$ \\
$0.346\pm0.038$ & $0.17\pm0.05$ & $0.26\pm0.04$ & $0.08\pm0.08$ \\
$0.422\pm0.038$ & $0.22\pm0.04$ & $0.12\pm0.05$ & $0.05\pm0.05$ \\
$0.499\pm0.038$ & $0.12\pm0.09$ & $0.08\pm0.07$ & $0.22\pm0.03$ \\
$0.576\pm0.038$ & $0.23\pm0.02$ & $0.08\pm0.06$ & $0.22\pm0.01$ \\
$0.653\pm0.038$ & $0.11\pm0.02$ & $0$ & $0.17\pm0.02$ \\
$0.730\pm0.038$ & $0.14\pm0.01$ & $0.03\pm0.03$ & $0.21\pm0.01$ \\
$0.806\pm0.038$ & $0.16\pm0.01$ & $0.13\pm0.01$ & $0.20\pm0.01$ \\
$0.883\pm0.038$ & $0.14\pm0.01$ & $0.16\pm0.01$ & $0.21\pm0.01$ \\
$0.960\pm0.038$ & $0.185\pm0.009$ & $0.152\pm0.009$ & $0.213\pm0.009$ \\
$1.039\pm0.038$ & $0.160\pm0.008$ & $0.13\pm0.01$ & $0.198\pm0.009$ \\
$1.114\pm0.038$ & $0.260\pm0.005$ & $0.11\pm0.01$ & $0.213\pm0.007$ \\
$1.190\pm0.038$ & $0.114\pm0.009$ & $0.13\pm0.01$ & $0.12\pm0.01$ \\
$1.267\pm0.038$ & $0.101\pm0.009$ & $0.11\pm0.02$ & $0.11\pm0.01$ \\
$1.344\pm0.038$ & $0.097\pm0.009$ & $0.10\pm0.01$ & $0.10\pm0.01$ \\
$1.421\pm0.038$ & $0.079\pm0.010$ & $0.09\pm0.01$ & $0.09\pm0.01$ \\
$1.498\pm0.038$ & $0.100\pm0.008$ & $0.11\pm0.01$ & $0.108\pm0.009$ \\
$1.575\pm0.038$ & $0.05\pm0.01$ & $0.06\pm0.02$ & $0.06\pm0.01$ \\
$1.651\pm0.038$ & $0.079\pm0.006$ & $0.078\pm0.007$ & $0.088\pm0.008$ \\
$1.728\pm0.038$ & $0.076\pm0.006$ & $0.073\pm0.007$ & $0.084\pm0.008$ \\
$1.805\pm0.038$ & $0.062\pm0.008$ & $0.062\pm0.007$ & $0.070\pm0.008$ \\
$1.882\pm0.038$ & $0.078\pm0.006$ & $0.078\pm0.006$ & $0.088\pm0.007$ \\
$1.959\pm0.038$ & $0.064\pm0.006$ & $0.063\pm0.006$ & $0.072\pm0.007$ \\
$2.035\pm0.038$ & $0.074\pm0.005$ & $0.072\pm0.005$ & $0.083\pm0.006$ \\
$2.112\pm0.038$ & $0.060\pm0.006$ & $0.059\pm0.006$ & $0.067\pm0.007$ \\
$2.189\pm0.038$ & $0.060\pm0.006$ & $0.059\pm0.006$ & $0.066\pm0.007$ \\
$2.266\pm0.038$ & $0.069\pm0.005$ & $0.068\pm0.005$ & $0.077\pm0.005$ \\
$2.343\pm0.038$ & $0.070\pm0.004$ & $0.086\pm0.006$ & $0.083\pm0.006$ \\
$2.419\pm0.038$ & $0.062\pm0.005$ & $0.076\pm0.007$ & $0.074\pm0.006$ \\
$2.496\pm0.038$ & $0.053\pm0.006$ & $0.065\pm0.006$ & $0.064\pm0.007$ \\
$2.573\pm0.038$ & $0.064\pm0.005$ & $0.078\pm0.005$ & $0.076\pm0.006$ \\
$2.650\pm0.038$ & $0.061\pm0.004$ & $0.074\pm0.005$ & $0.072\pm0.006$ \\
$2.727\pm0.038$ & $0.055\pm0.004$ & $0.068\pm0.006$ & $0.066\pm0.006$ \\
$2.803\pm0.038$ & $0.055\pm0.005$ & $0.067\pm0.005$ & $0.066\pm0.006$ \\
$2.880\pm0.038$ & $0.058\pm0.004$ & $0.072\pm0.006$ & $0.069\pm0.005$ \\
$2.957\pm0.038$ & $0.053\pm0.004$ & $0.065\pm0.005$ & $0.064\pm0.005$ \\
\hline
\end{tabular}
\end{center}
\label{tab:dens}
\end{minipage}
\end{table*}

\bibliographystyle{mnras_mwilliams} 
\bibliography{refs.bib}

% required because of bug in MN2e style file
% throws away figs otherwise
\clearpage

\end{document}